\documentclass[acmtog]{acmart}

\usepackage{booktabs} 

\usepackage[ruled]{algorithm2e} 

\SetAlFnt{\small}
\SetAlCapFnt{\small}
\SetAlCapNameFnt{\small}
\SetAlCapHSkip{0pt}

\acmJournal{TOG}

\begin{document}
\title{One-shot Embroidery Customization via Contrastive LoRA Modulation}

\author{Jun Ma}
\email{majun88818@163.com}
\affiliation{%
  \institution{Zhejiang Sci-Tech University and Style3D Research}
  \city{Hangzhou}
  \country{China}
}

\author{Qian He}
\email{heqianhailie@gmail.com}
\authornote{Project lead and corresponding author.}
\affiliation{%
  \institution{State Key Lab of CAD\&CG, Zhejiang University and Style3D Research}
  \city{Hangzhou}
  \country{China}
}

\author{Gaofeng He}
\email{hegaofeng@linctex.com}
\affiliation{%
  \institution{Style3D Research}
  \city{Hangzhou}
  \country{China}
}

\author{Huang Chen}
\email{chenhuang@linctex.com}
\affiliation{%
  \institution{Style3D Research}
  \city{Hangzhou}
  \country{China}
}

\author{Chen Liu}
\email{eric.liu@linctex.com}
\affiliation{%
  \institution{State Key Lab of CAD\&CG, Zhejiang University and Style3D Research}
  \city{Hangzhou}
  \country{China}
}

\author{Xiaogang Jin}
\email{jin@cad.zju.edu.cn}
\affiliation{%
  \institution{State Key Lab of CAD\&CG, Zhejiang University}
  \city{Hangzhou}
  \country{China}
}

\author{Huamin Wang}
\email{wanghmin@gmail.com}
\affiliation{%
  \institution{Style3D Research}
  \city{Hangzhou}
  \country{China}
}

\begin{abstract}
Diffusion models have significantly advanced image manipulation techniques, and their ability to generate photorealistic images is beginning to transform retail workflows, particularly in presale visualization. 
Beyond artistic style transfer, the capability to perform fine-grained visual feature transfer is becoming increasingly important. 
Embroidery is a textile art form characterized by intricate interplay of diverse stitch patterns and material properties, which poses unique challenges for existing style transfer methods. 
To explore the customization for such fine-grained features, we propose a novel contrastive learning framework that disentangles fine-grained style and content features with a single reference image, building on the classic concept of image analogy. 
We first construct an image pair to define the target style, and then adopt a similarity metric based on the decoupled representations of pretrained diffusion models for style-content separation. Subsequently, we propose a two-stage contrastive LoRA modulation technique to capture fine-grained style features. In the first stage, we iteratively update the whole LoRA and the selected style blocks to initially separate style from content. In the second stage, we design a contrastive learning strategy to further decouple style and content through self-knowledge distillation. Finally, we build an inference pipeline to handle image or text inputs with only the style blocks. To evaluate our method on fine-grained style transfer, we build a benchmark for embroidery customization. Our approach surpasses prior methods on this task and further demonstrates strong generalization to three additional domains: artistic style transfer, sketch colorization, and appearance transfer. Our project is available at: \href{https://style3d.github.io/embroidery_customization}{https://style3d.github.io/embroidery\_customization}.
\end{abstract}

%
%


%
%
\setcopyright{acmlicensed}
\acmJournal{TOG}
\acmYear{2025} \acmVolume{44} \acmNumber{6} \acmArticle{271} \acmMonth{12}\acmDOI{10.1145/3763290}


\begin{teaserfigure}
    \centering
    \includegraphics[width=\textwidth]{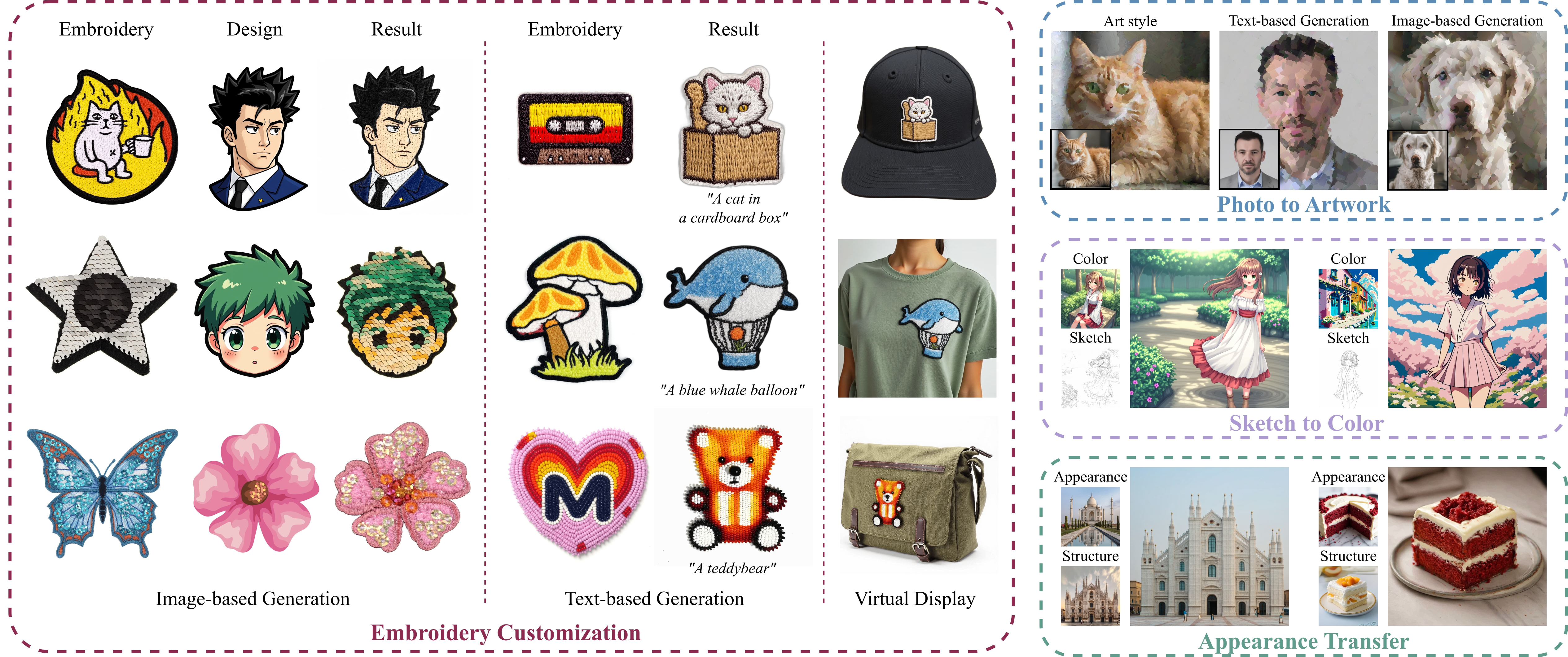}
    \caption{Given a reference embroidery image (Columns 1 and 4), our method is able to generate novel embroidery images (Columns 3 and 5) based on either image inputs (Column 2) or textual descriptions (Column 5). These outputs can be further integrated with ACE++~\cite{mao2025ace++} to produce decorative images for virtual display applications (Column 6). Furthermore, our approach demonstrates strong generalization across a range of visual attribute transfer tasks, including artistic style transfer (Row 1, last column), sketch colorization (Row 2, last column), and appearance transfer (Row 3, last column). Pink flower design (Row 3, Column 2) \textcopyright~ \href{https://www.vecteezy.com/vector-art/1931836-flower-pink-painting-vector-design}{Vecteezy}.}
    \label{fig:teaser}
\end{teaserfigure}

\maketitle

\section{Introduction}
\label{sec:intro}
\begin{figure}
    \centering
    \includegraphics[width=0.45\textwidth]{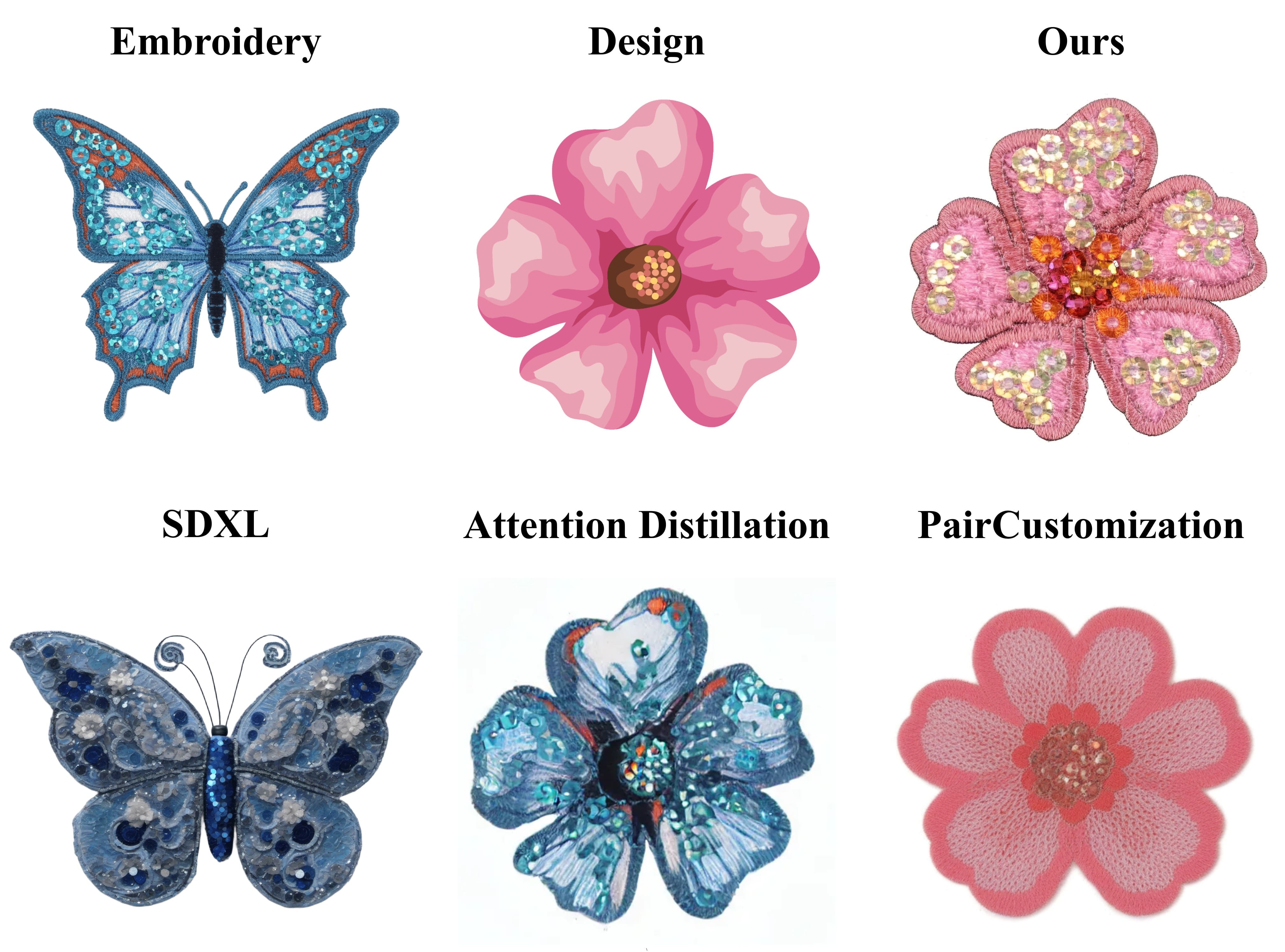}
    \caption{Challenges in embroidery customization. Given a reference embroidery, SDXL~\cite{podell2023sdxl} generates dissimilar style with merely text input "blue butterfly embroidery patch with sequins". Attention Distillation~\cite{zhou2025attention} and PairCustomization~\cite{jones2024customizing} both fail to capture sequin structure and yarns while maintainig the pink flower design, comparing to Ours. Pink flower design (Row 1, Column 2) \textcopyright~ \href{https://www.vecteezy.com/vector-art/1931836-flower-pink-painting-vector-design}{Vecteezy}.}
    \label{fig:intro}
\end{figure}

Visual attribute transfer~\cite{hertzmann2001image,efros2001image,liao2017visual} represents a fundamental challenge in image manipulation, involving the separation and recombination of style and content~\cite{tenenbaum1996separating}, and is revitalized by recent advances in diffusion-based generative models~\cite{rombach2022high,podell2023sdxl}, particularly in the domain of artistic style transfer~\cite{zhang2023inversion,wang2024instantstyle,zhou2025attention}. 
Beyond virtual display, diffusion models now generate high-quality, photorealistic images from customized instructions, often outperforming traditional 3D modeling and rendering. This capability is beginning to transform and even revolutionize retail workflows under the "sell it before you make it" paradigm~\cite{lin2025sell}, offering a novel approach to inventory challenges.
However, controllability over fine-grained structural elements such as embroidery or real-world textiles remains a key challenge.

Embroidery is an intricate textile art characterized by the structured arrangement of diverse yarns and materials, as shown in Fig.~\ref{fig:teaser}. Customization of embroidery styles poses unique challenges for existing methods in visual style transfer. To begin with, relying solely on pretrained models often fails to generalize to unseen embroideries~\cite{wang2024instantstyle,chung2024style}, while fine-tuning on large-scale datasets~\cite{qi2024deadiff,xing2024csgo} also proves ineffective due to data scarcity and complex intra-class variation in embroidery patterns. 
Furthermore, general style transfer often treats color as a key component of style~\cite{frenkel2025implicit}, whereas embroidery style focuses on high-frequency structural textures, largely independent of color, causing existing network block selection methods to struggle with separating embroidery style from its graphic design content. 
Additionally, other effective constraints for general style disentanglement can still have difficulty in capturing complex embroidery styles~\cite{jones2024customizing}, as shown in Fig.~\ref{fig:intro}.

To address these challenges, we propose a novel framework for fine-grained style customization, with embroidery as a representative case. Our main idea is to employ contrastive learning with a single reference image to achieve style–content disentanglement, which comprises three aspects: 
Firstly, we revisit the classic concept from image analogy~\cite{efros2001image} by constructing a single image pair to define a style, to reduce ambiguity and avoid inconsistency among different style images. 
Secondly, we adopt a metric to measure feature similarity within an image pair and cross image pairs, leveraging the decoupled representations~\cite{liao2017visual} of pretrained diffusion models. 
Finally, we design a contrastive LoRA~\cite{lorarepo} modulation technique named EmoLoRA, to first capture style features in selected blocks, and then further decouple style from content with self-knowledge distillation. 

Specifically, our framework comprises the following steps: pair-wise data construction, analysis on SDXL, contrastive LoRA learning and model inference. For a reference embroidery image, we leverage the prior knowledge in SD3 \cite{esser2024scaling} combined with ControlNets \cite{zhang2023adding}, to generate its graphic design image. Then we analyze SDXL by computing the cosine similarity between the constructed image pair in an inversion-reconstruction pipeline~\cite{garibi2024renoise}, using self-attention output features in each transformer block. Blocks with low feature similarity demonstrate closer correlation to the specified style, which is the main distinction between the image pair. 
After data preparation and model analysis, we design a two-stage contrastive LoRA learning strategy, to alleviate the overfitting and "content leakage" issue of the standard LoRA. 
In the first stage, we iteratively update the whole LoRA and the selected style blocks with supervision on the generated graphic design and the reference embroidery image, respectively. This modulation mechanism constrains the style into the selected blocks while leaves major content features to the other blocks. 
In the second stage, we use the trained LoRA to generate complementary data and apply contrastive learning on the noised latent feature space~\cite{dalva2024noiseclr}, to further decouple style and content in the style blocks. Finally, we build a model inference pipeline to handle image or text inputs with only the style blocks. 

Our main contributions are summarized as follows:
\begin{itemize}
    \item We propose a novel contrastive learning framework for fine-grained style customization, by constructing an image pair to define a style and designing a contrastive LoRA modulation technique to decouple style and content.
    \item We introduce a new task, one-shot embroidery customization, which poses unique challenges with intricate structural features, and conduct analysis to verify its potential in transforming real-world embroidery workflows.
    \item We outperform existing approaches in separating embroidery style from design content and exhibit strong generalization to three additional domains: artwork-photo, color-sketch, and appearance-structure.
\end{itemize}

\section{Related Work}
\label{sec:related}

\paragraph{Diffusion-based Image Synthesis}
Image synthesis has achieved tremendous progress with the rise of diffusion-based generative models~\cite{sohl2015deep, ho2020denoising, dhariwal2021diffusion, rombach2022high, peebles2023scalable}. Harnessing the generative power of pre-trained text-to-image models, various applications in personalization / customization~\cite{zhang2023prospect, kumari2023multi, tang2024realfill, tewel2023key} are developed. Given a small image set in a new concept, Dreambooth~\cite{ruiz2023dreambooth} adapts the model via finetuning, while TI~\cite{gal2022image} finds the embeddings in the textual feature space. LoRA is a PEFT method~\cite{houlsby2019parameter} to adapt large language~\cite{hu2021lora} or vision~\cite{lorarepo} models to downstream tasks. 
Recent works~\cite{zhang2023adding, mou2024t2i, ye2023ip} propose plug-and-play adapters that enable controllable image generation by modulating the generative process without retraining the base model.
To handle image editing and translating~\cite{nichol2021glide, kawar2023imagic, brooks2023instructpix2pix, valevski2023unitune}, SDEdit~\cite{meng2021sdedit} first adds noise to the input image and then denoises it through the SDE prior, while~\cite{hertz2022prompt, tumanyan2023plug} adopt an inversion~\cite{song2020denoising, mokady2023null} pipeline and attention feature manipulation~\cite{liu2024towards}. Due to the limited expressiveness of text for fine-grained spacial features, we mainly explore LoRA-based methods to leverage the representation capacity of pretrained diffusion models, and propose to further decouple these features for style customization during finetuning.

\paragraph{Visual Attribute Transfer}
Visual attribute transfer~\cite{hertzmann2001image, efros2001image} aims to transform an image to adopt the style of another, encompassing elements such as color, texture, local structures, and artistic style. 
Neural style transfer approaches have evolved from early CNN-based frameworks~\cite{gatys2016image, johnson2016perceptual, huang2017arbitrary, li2017universal, park2019arbitrary, zhang2022domain}, to adversarial learning with GANs~\cite{karras2019style, karras2020analyzing, goodfellow2014generative, isola2017image, zhu2017unpaired, park2020contrastive}, and more recently to transformer architectures leveraging self-attention for global context~\cite{wu2021styleformer}. 
The rapid development of text-to-image diffusion models has also sparked their adaptation to style transfer tasks, including exploring the textual feature space \cite{yang2023zero, zhang2023inversion, qi2024deadiff, li2025styletokenizer}, manipulating the attention features \cite{chung2024style, deng2023z, hertz2024style}, leveraging plug-and-play adapters~\cite{wang2024instantstyle}, finetuning LoRAs \cite{shah2025ziplora, frenkel2025implicit, jones2024customizing}, and formulating the problem using stochastic optimal control~\cite{rout2024rb} with an existing style descriptor~\cite{somepalli2024measuring}. 
However, these methods fall short when handling intricate structural styles such as embroidery, where color serves as content rather than style, and high-frequency structural textures, commonly neglected in artistic style transfer, play a central role in characterizing style. A detailed discussion is provided in Sec.~\ref{sec:comparisons}. 
Sketch colorization~\cite{yan2025image, zhang2021line, li2022eliminating} and appearance transfer~\cite{alaluf2024cross, tumanyan2022splicing, kwon2022diffusion} are slightly different problems than artistic style transfer, as they have different definition of style and content, and potentially involve semantic correspondence between reference and target images. 
Attention Distillation~\cite{zhou2025attention} transfers style or appearance using distillation loss on pretrained attention features, yet still has difficulty in separating structural embroidery styles from color content, as presented in Fig.~\ref{fig:intro}. Furthermore, diffusion-based image analogy frameworks propose a more generic approach for visual attribute transfer. DIA~\cite{vsubrtova2023diffusion} focuses on high-level semantics and represents $A:A'$ in CLIP embedding space, while Analogist~\cite{gu2024analogist} uses GPT-4V~\cite{achiam2023gpt} to reason the analogy $A:A'::B:B'$ and relies on textual descriptions to capture style, both are limited in capturing fine-grained styles like embroidery. 
In this work, we propose a novel framework to capture intricate structural styles at a fine-grained level, thereby addressing challenges more closely aligned with real-world applications.

\begin{figure*}[t]
    \centering
    \includegraphics[width=\textwidth]{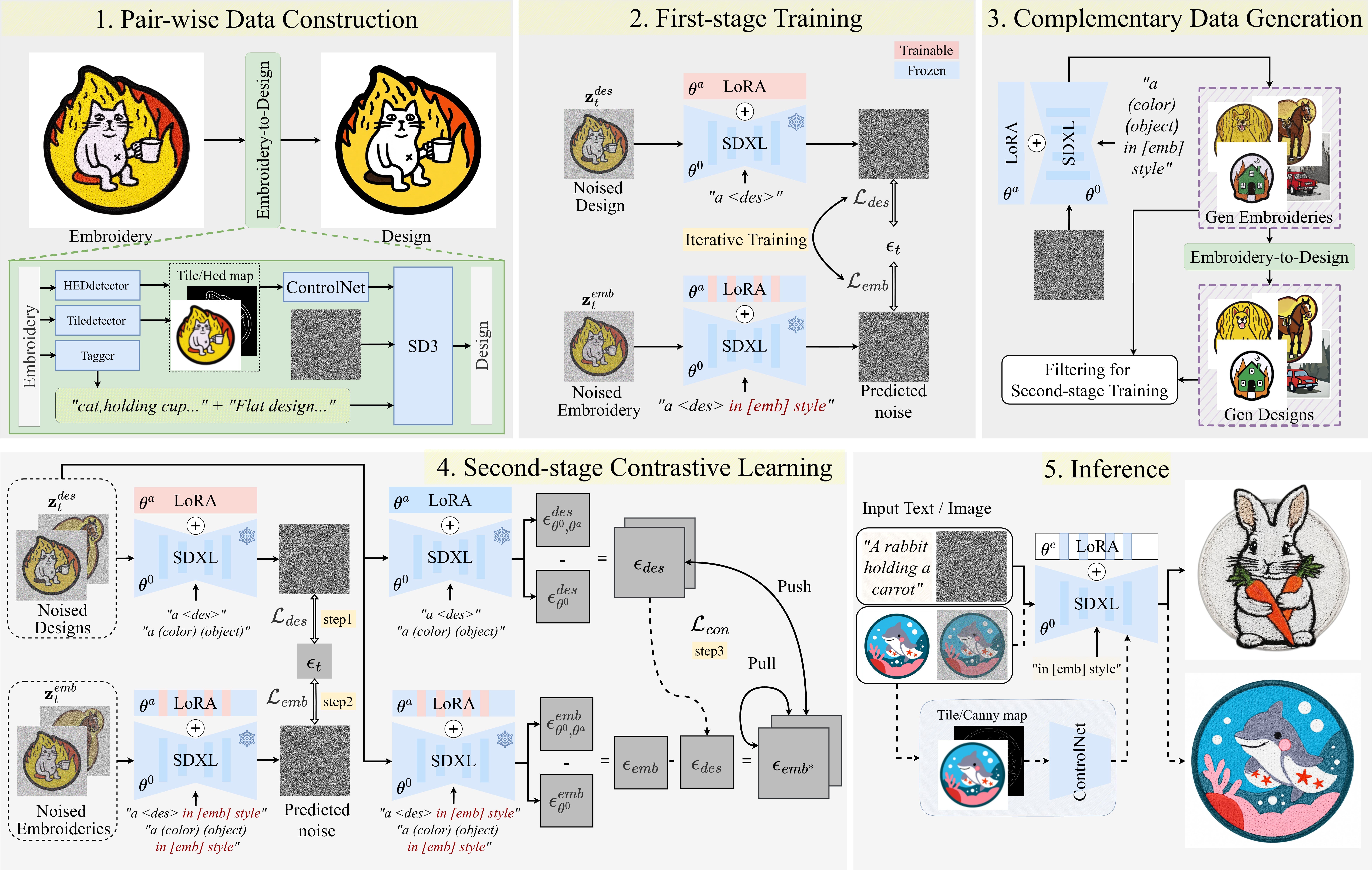}
    \caption{An overview of our framework. (1) Pair-wise data construction: We build a pipeline to process the reference embroidery into its graphic design; (2) First-stage training: Given the embroidery-design pair, we iteratively train a LoRA $\theta^a$ to initially decouple style and content; (3) Complementary data generation: We generate more embroidery-design pairs with the trained LoRA from the first stage; (4) Second-stage contrastive learning: We conduct contrastive learning to further decouple style and content in embroidery LoRA blocks $\theta^e$ using the reference and a generated emb-des pair; (5) Model inference: We use the trained embroidery LoRA $\theta^e$ to conduct text/image-based customization.}
    \label{fig:method}
\end{figure*}

\paragraph{Embroidery Synthesis} 
Embroidery is a decorative fabric art form \cite{nichols2012encyclopedia, pile2018fashion} that can take on various styles, each exhibiting distinct visual characteristics due to the use of different yarns and stitches, as well as materials such as pearls, beads, and sequins (Fig.~\ref{fig:teaser} and~\ref{fig:img2img1}). 
For example, chenille embroidery creates a fuzzy, textured surface with looped yarn (Row 2, Column 4 in Fig.~\ref{fig:teaser}), while sequin embroidery incorporates reflective discs secured by stitches for added sparkle (Row 3, Column 1 in Fig.~\ref{fig:teaser}). 
Contemporary embroidery design heavily relies on specialized CAD software (e.g., Wilcom EmbroideryStudio\footnote{https://wilcom.com/}) that translates digital artwork into machine-readable stitch instructions, but the process still involves extensive manual operation and often lacks fully satisfactory visualizations. 
Automated embroidery synthesis has continued to attract interest as a specialized but intriguing topic in computer graphics and digital fabrication. 
Early approaches model flat embroidery in 3D by incorporating three fundamental stitch types—long-short, satin, and stem/edge stitches—into geometric representations \cite{chen2012embroidery, cui2017image}. Subsequent work focuses on simulating random-needle embroidery using vector fields or stitch primitives, followed by multilayer rendering techniques to enhance visual fidelity \cite{yang2012image, yang2018paint, ma2022multilayered}.
To improve rendering realism, intrinsic image decomposition is introduced to preserve the illumination of input photographs during the synthesis process \cite{shen2017illumination}. The manual specification of stitch types is further alleviated by segmenting input images into subregions, assigning appropriate stitch categories to each segment, and synthesizing embroidery textures via UV mapping \cite{guan2021automatic}. More recent approaches leverage deep neural networks and adversarial learning to directly generate embroidery-like imagery \cite{ye2021towards, yang2022unsupervised}.
Empowered by supervised deep learning, stitch types in segmented subregions can be explicitly classified using annotated datasets \cite{hu2024msembgan}. Beyond visual realism, recent efforts have also explored the generation of machine-fabricable embroidery patterns. For example, Liu et al. \shortcite{zhenyuan2023embroidery}  propose a method that uses user-defined directional cues and vector field analysis to derive continuous streamlines suitable for machine stitching.
In contrast to prior methods, our approach takes as input a reference embroidery image and a natural language prompt, and generates a customized embroidery image. It supports a wide variety of stitches and materials without relying on explicit stitch-type labeling or manual annotations.

\section{Method}
\label{sec:method}
In this section, we introduce our contrastive learning framework for fine-grained style customization, using embroidery as a representative case.
Given a reference embroidery image $I$, our objective is to generate embroidery images that replicate the same style, encompassing stitches, yarns, accessories, and other prominent structural features. 
To enable contrastive learning with a single reference image, we first introduce a pair-wise data construction module for style definition in Sec.~\ref{sec:data_pair} and a similarity metric for identifying style features in Sec.~\ref{sec:sdxl_analysis}. Building on this foundation, we then present our two-stage contrastive LoRA learning strategy for style-content disentanglement in Sec.~\ref{sec:contrastive}. 
Finally, we describe our model inference in Sec.~\ref{sec:infer}. An overview of our framework is in Fig.~\ref{fig:method}.

\begin{figure*}
    \centering
    \includegraphics[width=\textwidth]{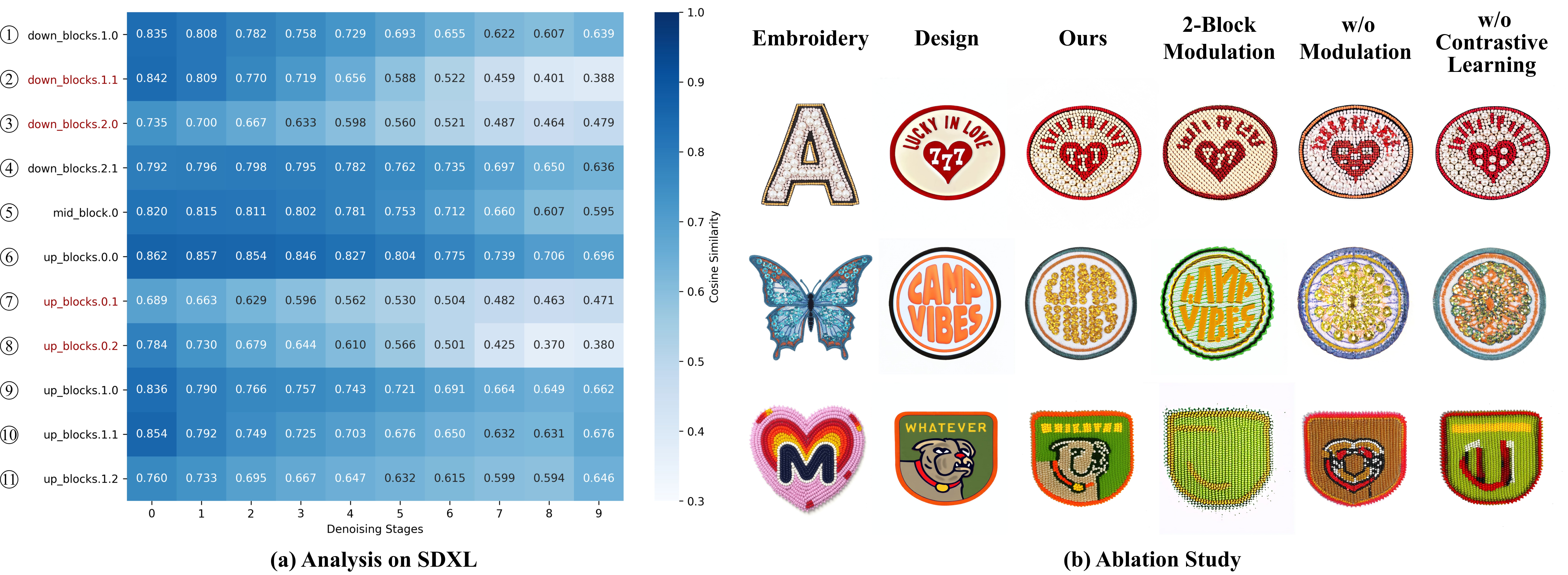}
    \caption{Base model analysis and ablation study. (a) Average cosine similarity of self-attention outputs between all paired embroidery and design images in our reference set. A cosine similarity of 1 indicates that the block produces nearly identical features for an embroidery-design pair, implying that it primarily represents non-embroidery content. (b) Ablation study on three reference embroidery examples.}
    \label{fig:analysis_ablation}
\end{figure*}

\subsection{Pair-wise Data Construction}
\label{sec:data_pair}
Given a single reference embroidery image, our goal is to disentangle style from content to provide supervision for contrastive learning. While embroidery style is abstract and difficult to capture explicitly, design content is comparatively easier to represent. To this end, we construct a data pipeline that generates a corresponding graphic design image, thereby defining style through a data pair in the spirit of image analogy~\cite{efros2001image}. 
In this embroidery-to-design module, we adopt the text-to-image pipeline of SD3~\cite{esser2024scaling} with ControlNet~\cite{zhang2023adding}. Specifically, we utilize the HEDdetector~\cite{xie2015holistically} to detect rough edges of the embroidery image, and then send the edge image to ControlNet-Canny. In this way, the design structure is well preserved, while the embroidery stitches are effectively removed. To preserve color fidelity and enhance generation quality, we send the blurred embroidery image into a ControlNet-Tile branch. Moreover, we use WD14~\cite{smilingwolf2023wdconvnexttagger} to generate captions as the prompt with "\emph{flat design, vector graphic design, digital design, cartoon design, clean lines, uniform color blocks, smooth surface, high quality}" and thus achieve better preservation of the design content. Empirically, we find that using the pipeline with SD3 yields better results than SDXL, which is probably due to differences in their pretraining datasets. While our embroidery-to-design module may not generalize directly to other styles, the underlying concept of pair-wise data construction can be adapted through alternative means.

\subsection{Analysis on SDXL}
\label{sec:sdxl_analysis} 
Inspired by B-LoRA~\cite{frenkel2025implicit}, we leverage the decoupled representations of pretrained diffusion models and employ different LoRA blocks to separate style and content. 
Unlike B-LoRA, however, embroidery style cannot be captured by a single block entangled with color information, as it is largely independent of color, nor can a single content block fully reconstruct the detailed design. 
Building on prior findings~\cite{tumanyan2023plug, liu2024towards} that self-attention features in the UNet of diffusion models encode spatial structure, including high-frequency details, we introduce a similarity metric to guide the selection of network blocks most suitable for capturing a specified style. 
We adopt SDXL as our base model for its higher resolution, improved visual fidelity, and more naturally decoupled attention features compared to SD3. 
In this work, we employ ReNoise~\cite{garibi2024renoise}, an inversion technique that achieves higher reconstruction quality than DDIM~\cite{song2020denoising}, and leverage the output features of each self-attention layer in the image reconstruction process to compare differences between each embroidery-design pair. 
Let $\mathbf{F}_i$ denote the input feature to a self-attention layer, and $f_i^q(\cdot)$, $f_i^k(\cdot)$, $f_i^v(\cdot)$ and $f_i^o(\cdot)$ be the projection layers for query, key, value and output, respectively. We have $\mathbf{Q}_i = f_i^q(\mathbf{F}_i)$, $\mathbf{K}_i = f_i^k(\mathbf{F}_i)$, $\mathbf{V}_i = f_i^v(\mathbf{F}_i)$, and $d_k$ as the dimensionality of query and key. The output self-attention feature is: 
\begin{equation}
    \mathbf{F}_i^o = f_i^o(Softmax(\frac{\mathbf{Q}_i \mathbf{K}_i^T}{\sqrt{d_k}})\mathbf{V}_i).
\end{equation}

For each block, we compute the average cosine similarity of all $\mathbf{F}_i^o$ between an embroidery-design pair. Since the whole generation process consists of 50 steps, we divide them into 10 sections and compute the average. Empirically, we find the similarity matrix for the collected reference embroidery set share the same relative scale and thus adopt their average as shown in Fig.~\ref{fig:analysis_ablation} (a), where a cosine similarity of 1 indicates that the block produces nearly identical features for an embroidery-design pair, implying that it primarily represents non-embroidery content. 
We notice that \emph{down\_blocks 1.1, 2.0} and \emph{up\_blocks 0.1, 0.2} show more correlation to embroidery features as they differ more between embroidery and design. The difference is more significant in later stages of the denoising process, mainly because embroidery style is more related to delicate low-level image features than rough high-level semantic features.

\subsection{Contrastive LoRA Learning}
\label{sec:contrastive}
With the constructed embroidery-design pairs and network feature analysis, we now present a two-stage contrastive LoRA learning strategy named EmoLoRA, as in Alg.~\ref{alg:emolora}, that captures embroidery style from a single reference while alleviating overfitting of standard LoRA~\cite{lorarepo}. In this work, LoRA is applied to the attention layers of the UNet backbone. 
During the first stage, we train EmoLoRA to roughly decouple embroidery and design through a block modulation mechanism. In the second stage, we use the trained EmoLoRA to generate embroidery images with preset prompts for more supervision signals, and then adopt contrastive learning to further enhance the decoupling of embroidery and design.

\paragraph{LoRA Block Modulation}
For our EmoLoRA, we separate the four blocks discussed in Sec.~\ref{sec:sdxl_analysis} to capture embroidery style while using the whole LoRA to recover the design content. 
Given an embroidery-design pair, we set the prompt for the embroidery image as "\emph{a <des> in [emb] style}", and the prompt for the design as "\emph{a <des>}". During training, we update EmoLoRA weights in a two-step iterative manner. In Step 1, we input "\emph{a <des>}" into all blocks of the SDXL base model $\theta^0$ and EmoLoRA $\theta^a$, and train $\theta^a$ with $\mathcal{L}_{des}(\theta^a)$. In Step 2, we input "\emph{a <des> in [emb] style}" into the four embroidery blocks of SDXL and EmoLoRA, and "\emph{a <des>}" into all other blocks, and only update the four embroidery blocks $\theta^e$ of EmoLoRA with $\mathcal{L}_{emb}(\theta^e)$. $\mathbf{z}_t^{des}$ and $\mathbf{z}_t^{emb}$ denote the noised image features in latent space~\cite{kingma2013auto, esser2021taming}, $\mathbf{c}_{des}$ and $\mathbf{c}_{emb}$ are the encoded text features~\cite{Radford2021LearningTV}. Note that $\theta^e$ is a subset of $\theta^a$. We define $\mathcal{L}_{des}(\theta^a)$ and $\mathcal{L}_{emb}(\theta^e)$ as:
\begin{equation}
\label{eq:l_des}
	\mathcal{L}_{des}(\theta^a) = || \mathbf{\epsilon}_t - \mathbf{\epsilon}_{\theta^0,\theta^a}(\mathbf{z}_t^{des}, t, \mathbf{c}_{des})||^2_2,
\end{equation}
\begin{equation}
\label{eq:l_emb}
	\mathcal{L}_{emb}(\theta^e) = || \mathbf{\epsilon}_t - \mathbf{\epsilon}_{\theta^0,\theta^a}(\mathbf{z}_t^{emb}, t, \mathbf{c}_{emb})||^2_2.
\end{equation}

After the iterative training process, the embroidery style is encapsulated solely in $\theta^e$ and decoupled from the main content, as only $\theta^e$ is updated during Step 2.
However, $\theta^e$ also contains some content information learned from Step 1, which is unable to avoid as the other blocks alone cannot recover the whole content image. 
Consequently, the generated images may retain color from the reference embroidery and have suboptimal fusion with the new content due to entanglement between the style and its original content, as shown in Fig.~\ref{fig:analysis_ablation} (b) \textbf{w/o Contrastive Learning}. To further alleviate this problem, we adopt a second stage with contrastive learning.

\begin{algorithm}[t]
\caption{Two-stage Contrastive EmoLoRA Learning}
\label{alg:emolora}
\KwIn{Constructed embroidery-design image pair $(I_{emb}, I_{des})$, text pair ("a <des> in [emb] style", "a <des>"), SDXL base model $\theta^0$, EmoLoRA $\theta^a$, selected embroidery blocks $\theta^e$ of EmoLoRA, learning rates $\eta_1$ and $\eta_2$}
\KwOut{Trained EmoLoRA $\theta^a$ with embroidery blocks $\theta^e$}

\BlankLine
\textbf{Stage 1: LoRA Block Modulation}\\
\For{each iteration}{
    \BlankLine
    Step 1, Update $\theta^a$ with $I_{des}$: 
    $\theta^a \leftarrow \theta^a - \eta_1 \nabla_{\theta^a} \mathcal{L}_{des}$\;

    \BlankLine
    Step 2, Update $\theta^e$ with $I_{emb}$: 
    $\theta^e \leftarrow \theta^e - \eta_1 \nabla_{\theta^e} \mathcal{L}_{emb}$\;
}

\BlankLine
\textbf{Complementary Data Generation:} \\
\Indp
Prompt: "a (color) (object) in [emb] style" $\times N$\;
Generate embroidery images and select $\lceil N / 2 \rceil$\;
Obtain design images and select $\lceil N / 4 \rceil$\;
\Indm

\BlankLine
\textbf{Stage 2: Contrastive Learning}\\
\For{each iteration}{
    \BlankLine
    Sample a generated pair $(I_{emb}^{gen}, I_{des}^{gen})$\;

    \BlankLine
    Step 1, Update $\theta^a$ with $I_{des}$ and $I_{des}^{gen}$: 
    $\theta^a \leftarrow \theta^a - \eta_1 \nabla_{\theta^a} \mathcal{L}_{des}$\;

    \BlankLine
    Step 2, Update $\theta^e$ with $I_{emb}$ and $I_{emb}^{gen}$: 
    $\theta^e \leftarrow \theta^e - \eta_1 \nabla_{\theta^e} \mathcal{L}_{emb}$\;

    \BlankLine
    Step 3, Update $\theta^e$ with $(I_{emb}, I_{des})$ and $(I_{emb}^{gen}, I_{des}^{gen})$: \\
    \Indp
    $\theta^e \leftarrow \theta^e - \eta_2 \nabla_{\theta^e} \mathcal{L}_{con}$\;
    \Indm
}
\end{algorithm}

\paragraph{Complementary Data Generation}
Before applying contrastive learning, we generate more data using the trained EmoLoRA from the first stage. 
We generate new embroidery images with a predefined set of prompts in "\emph{a (color) (object) in [emb] style}", covering various color-object combinations and blending the prior knowledge in SDXL with the learned embroidery style.
One example is "\emph{a yellow dog in [emb] style}", while the list of all $N$ prompts is in our supplementary and we set $N$ to 10 in this paper. We then compute the average cosine similarity of each generated image to the reference image using their self-attention output features, as in Sec.~\ref{sec:sdxl_analysis}, to measure the style similarity. Since embroidery features are more salient in later generation stages, we only use features from stages 5-9. Then we rank the generated images w.r.t. their average similarity and select the top half $\lceil N/2 \rceil$ for better embroidery quality, and use the embroidery-to-design pipeline in Sec.~\ref{sec:data_pair} to obtain their corresponding design images. Finally, we want to remove images with content that is too similar to the reference. So we compute and rank the average cosine similarity among the design images similar to embroidery images as before, and choose the most dissimilar half $\lceil N / 4 \rceil$ to be our final complementary data.

\paragraph{Contrastive Learning}
With the embroidery-design pairs from the initial reference and complementary generation, we now apply contrastive training. The main objective is to pull the embroidery features shared by different image pairs together, and to push away the embroidery features from the content features. Inspired by NoiseCLR~\cite{dalva2024noiseclr}, we conduct contrastive learning in the noised latent feature space. We obtain the design content features $\mathbf{\epsilon}_{des}$ by subtracting base model prediction from base model with EmoLoRA prediction, given noised design image features $\mathbf{z}_t^{des}$ and encoded text features $\mathbf{c}_{des}$ at timestep $t$. Similarly, we can obtain the embroidery image features $\mathbf{\epsilon}_{emb}$, while with noised design image features $\mathbf{z}_t^{des}$ but embroidery prompt features $\mathbf{c}_{emb}$. In this way, we are able to separate the learned knowledge in EmoLoRA $\theta^a$ that is triggered by "a <des>" or "a <des> in [emb] style". The formulation is as follows:
\begin{equation}
	\mathbf{\epsilon}_{des} = \mathbf{\epsilon}_{\theta^0, \theta^a}(\mathbf{z}_t^{des}, t, \mathbf{c}_{des}) - \mathbf{\epsilon}_{\theta^0}(\mathbf{z}_t^{des}, t,  \mathbf{c}_{des}),
\end{equation}
\begin{equation}
	\mathbf{\epsilon}_{emb} = \mathbf{\epsilon}_{\theta^0, \theta^a}(\mathbf{z}_t^{des}, t, \mathbf{c}_{emb}) - \mathbf{\epsilon}_{\theta^0}(\mathbf{z}_t^{des}, t, \mathbf{c}_{emb}),
\end{equation}
\begin{equation}
	\mathbf{\epsilon}_{emb^*} = \mathbf{\epsilon}_{emb} - \mathbf{\epsilon}_{des}.
\end{equation}

Note that $\mathbf{\epsilon}_{emb}$ also contains design content information, and should not be pushed away from $\mathbf{\epsilon}_{des}$. Therefore, we use $\mathbf{\epsilon}_{emb^*}$ to represent the subtracted embroidery features alone and push it away from the content features. We construct training batches, where each batch consists of the reference embroidery-design pair and a generated pair. For each batch, the final contrastive loss, denoted as $\mathcal{L}_{con}(\theta^e)$, is defined as follows:
\begin{equation}
	\mathcal{L}_{con}(\theta^e) = -\log \frac{\exp(s(\mathbf{\epsilon}_{emb^*}^{ref}, \mathbf{\epsilon}_{emb^*}^{gen}))}{\exp(s(\mathbf{\epsilon}_{emb^*}^{ref}, \mathbf{\epsilon}_{des}^{gen})) + \exp(s(\mathbf{\epsilon}_{des}^{ref}, \mathbf{\epsilon}_{emb^*}^{gen}))}.
\end{equation}

Here, we set the temperature $\tau$ to $1$ and omit it for simplicity, and $s(\cdot, \cdot)$ denotes cosine similarity. 
The complementary embroidery images are from EmoLoRA generation and can have a good initial $\mathbf{\epsilon}_{emb}^{gen}$, while the design images are from the embroidery-to-design pipeline and make $\mathbf{\epsilon}_{des}^{gen}$ unreasonable. To deal with this, we adopt a three-step iterative optimization. In Steps 1 and 2, we update $\theta^a$ and $\theta^e$ in Eqs.~\ref{eq:l_des}-\ref{eq:l_emb} as in the first stage, but on both the reference pair and the generated pair. 
In Step 3, we update $\theta^e$ with the contrastive loss $\mathcal{L}_{con}(\theta^e)$.

\subsection{Model Inference}
\label{sec:infer}
After the two-stage training of EmoLoRA, we apply model inference with image or text inputs, as in Alg.~\ref{alg:inference}. For both settings, we only use the four embroidery blocks $\theta^e$ to update SDXL base model $\theta^0$. For text inputs, which should include "\emph{in [emb] style}", the model performs standard text-to-image synthesis. For image inputs, we adopt SDEdit~\cite{meng2021sdedit} to first add noise to the input image, and then utilize the updated model to perform denoising under the guidance of text prompt "\emph{in [emb] style}".

Similar to pair-wise data construction, we employ ControlNets to maintain the content from the input image, according to the style type. For styles such as flat embroidery, the boundaries between an image pair should be accurately aligned, we employ ControlNet-Tile and ControlNet-Canny, followed by a color correction module to enhance consistency with the input design. In this module, we first transfer the generated embroidery image to LAB space, then replace its A and B channels with the corresponding channels from the input design, and finally transfer the embroidery image back to RGB space. 
However, for embroideries with beads or sequins, we disable ControlNet-Canny and the color correction module to allow necessary modifications along the boundaries. Similar principles can be extended to other styles.

\section{Experiment}
\label{sec:exp}
We build a benchmark on embroidery customization to evaluate our method against prior art for fine-grained style customization, with ablation study to verify the efficacy of each component. 
In applications with customized embroideries, we explore the potential for transforming traditional embroidery workflows. 
Additionally, we extend our method to three additional style transfer tasks to illustrate its capability in decoupling style and content. 
For more implementation details, results, and discussions, please refer to the supplementary material.

\begin{algorithm}[t]
\caption{EmoLoRA Inference}
\label{alg:inference}
\KwIn{Input text prompt $p$ or design image $\hat{I}_{des}$, SDXL base model $\theta^0$, trained embroidery blocks $\theta^e$, ControlNet-Tile\&Canny}
\KwOut{Embroidery image $\hat{I}_{emb}$}

\BlankLine
Update SDXL: $\theta \leftarrow \theta^0 + \theta^e$\;
Update prompt: $p \leftarrow p \  + \ $"in [emb] style"\;

\BlankLine
\If{$\hat{I}_{des}$ is not empty}{
    \BlankLine
    Employ ControlNet-Tile\;
    \BlankLine
    \If{Strict boundary alignment}{
    \BlankLine
    Employ ControlNet-Canny and color-correction\;
    }
}

\BlankLine
Generate image: $\hat{I}_{emb}$.

\end{algorithm}

\subsection{Embroidery Dataset and Metrics}
\label{sec:eval}

\paragraph{Dataset}
We follow style transfer benchmarks~\cite{deng2022stytr2,chung2024style,deng2023z} and build a dataset comprising 30 reference embroidery images and 50 test graphic design images. 
The reference set contains embroidery styles featuring various stitches and materials, including flat stich, towel stitch, beans, sequins, and more. For the test set, we use our embroidery-to-design module to generate the graphic design images with 50 additional embroidery images in any style, ensuring the test images are compatible with embroidery production. Additionally, we preset 20 text prompts for text-based generation. For each reference image, we evaluate the method across all test images and prompts.

\paragraph{Metrics}
For image-based customization, we adopt LPIPS~\cite{zhang2018perceptual} and Histogram Loss~\cite{afifi2021histogan, chung2024style} to assess the preservation of design content and color. To evaluate embroidery style, we propose a metric named High-Frequency Ratio Difference (HFRD). Existing feature extractors such as VGG or CLIP mainly captures color, layout, or semantic features, while embroidery styles emphasize high-frequency structural textures, which makes existing metrics unsuitable for embroidery style evaluation. We propose HFRD to compute the absolute difference of the high-frequency energy ratio between the generated embroidery image and the reference. 
For text-based generation, we compute the CLIP-Score~\cite{Radford2021LearningTV} between the generated image and the text prompt, to evaluate the level of semantic compliance. Additionally, we adopt Histogram Loss to assess the color difference between the generated embroidery and the reference, where a higher score means less similar in color and therefore better decoupling from reference content. 
The details and limitations of the metrics are discussed in the supplementary material, and we provide user studies in Sec.~\ref{sec:user_study} to strengthen quantitative comparisons with different style transfer methods and our ablation variants.

\begin{figure*}
    \centering
    \includegraphics[width=\textwidth]{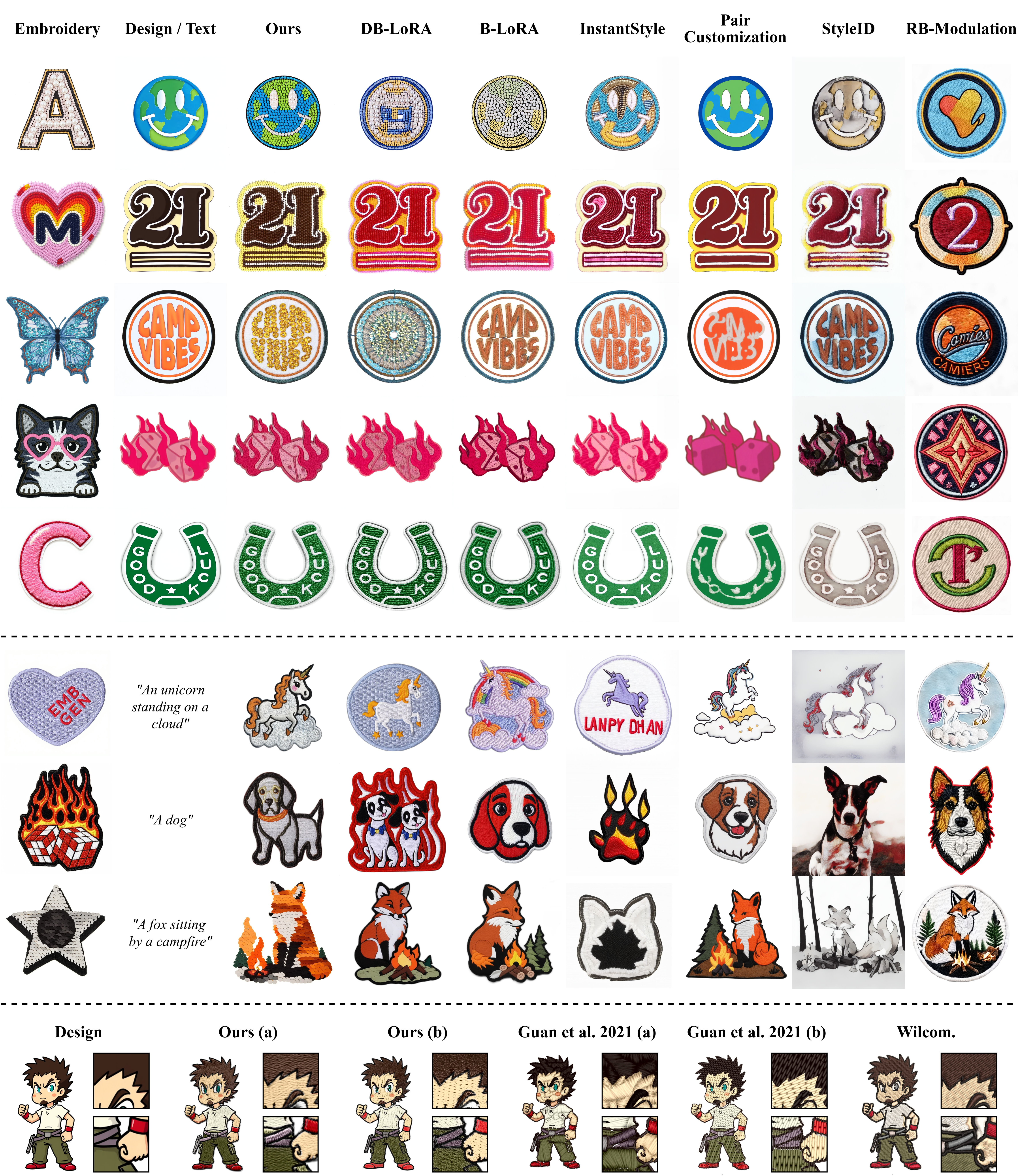}
    \caption{
    Comparison on one-shot embroidery customization using image/text inputs. The last row shows comparisons with two embroidery synthesis methods. \textbf{Ours (a)} and \textbf{Ours (b)} denote results using different reference embroideries. Please zoom in to examine detailed textures.
    }
    \label{fig:img2img1}
\end{figure*}

\begin{table*}[t]
    \centering
    \caption{Quantitative comparisons with style transfer methods. For image-based generation, we evaluate embroidery style quality (HFRD), design content preservation (LPIPS), and design color consistency (Histogram Loss). For text-based generation, we assess textual compliance (CLIP-Score) and reference color resemblance (Histogram Loss). The best results are highlighted in bold, and the second-best are underlined.}
    \scalebox{0.9}{
    \begin{tabular}{@{}lcccccccc@{}}
        \toprule
        Metric & Ours & DB-LoRA & B-LoRA & InstantStyle & PairCustomization & StyleID & RB-Modulation \\
        \midrule
        HFRD $\downarrow$ (embroidery style) & \textbf{6.50} $\pm$ 3.14 & 8.15 $\pm$ 4.22 & \underline{6.63} $\pm$ 2.25 & 12.41 $\pm$ 5.76 & 12.48 $\pm$ 5.17 & 21.66 $\pm$ 7.16 & 8.10 $\pm$ 3.97\\
        LPIPS $\downarrow$ (design content) & \underline{14.37} $\pm$ 9.66 & 14.54 $\pm$ 10.52 & 14.92 $\pm$ 8.03 & \textbf{7.72} $\pm$ 7.47 & 22.14 $\pm$ 2.63 & 21.96 $\pm$ 2.59 & 65.18 $\pm$ 1.65\\
        Histogram Loss $\downarrow$ (design color) & \textbf{26.59} $\pm$ 9.55 & \underline{28.62} $\pm$ 8.52 & 30.57 $\pm$ 7.97 & 32.23 $\pm$ 7.38 & 43.99 $\pm$ 1.99 & 45.75 $\pm$ 4.12 & 48.87 $\pm$ 1.61\\
        \midrule
        CLIP-Score $\uparrow$ (text semantics) & \underline{32.23} $\pm$ 0.61 & 30.94 $\pm$ 0.66 & 31.84 $\pm$ 0.33 & 25.14 $\pm$ 2.09 & \textbf{32.47} $\pm$ 0.23 & 30.31 $\pm$ 0.92 & 30.04 $\pm$ 0.37\\
        Histogram Loss $\uparrow$ (reference color) & \textbf{51.32} $\pm$ 6.82 & 43.89 $\pm$ 7.21 & 42.70 $\pm$ 6.09 & 33.64 $\pm$ 11.19 & \underline{50.49} $\pm$ 5.82 & 34.13 $\pm$ 9.89 & 35.48 $\pm$ 7.66\\
        \bottomrule
    \end{tabular}
    }
\label{tab:combined_results}
\end{table*}

\subsection{Comparison on Embroidery Customization}
\label{sec:comparisons}
Embroidery customization is a unique problem in contrast to general style transfer, as it redefines style and content. Elements like color must now be preserved as content rather than transferred as style, while high-frequency structural textures, often overlooked in artistic style transfer, become central to defining style. 
For embroidery customization, we focus on embroidery style similarity, design content preservation (for image inputs), decoupling of style from color and semantics (for text inputs). Qualitative results are in Fig.~\ref{fig:img2img1} and quantitative results are in Tab.~\ref{tab:combined_results}.

We first compare with six prior methods in style transfer. 
On nine different embroidery styles in Fig.~\ref{fig:img2img1}, our method exhibits high quality in fusion of the reference style and input image/text content. 
For DB-LoRA~\cite{lorarepo} and B-LoRA~\cite{frenkel2025implicit}, we use their training approach but with our inference pipeline, to fully evaluate the style-content decoupling capability of our EmoLoRA. 
DB-LoRA can capture complete pearls or beads, while fails to fuse these structures with input content (Row 1, 3, 8) due to entanglement of style and reference content, and maintains the reference color or layout (Row 2, 6, 7, 9). 
B-LoRA uses a single \emph{up\_blocks.0.1} to capture style, possessing limited power in capturing embroidery structures and still presenting entanglement with the reference color. 
InstantStyle~\cite{wang2024instantstyle} injects reference features via pretrained IP-Adapter into the cross-attention of \emph{up\_blocks.0.1}, and thus captures even less embroidery features than B-LoRA. 
PairCustomization~\cite{jones2024customizing} adopts two LoRAs with orthogonal constraints to disentangle style and content, while freezing orthogonal matrices and training only one low-rank matrix also fails to capture complex embroidery textures. 
StyleID~\cite{chung2024style} blends style and content latent from DDIM inversion and leverages self-attention in later blocks, but produces blurry results due to entangled style and content features in pretrained self-attention. 
RB-Modulation~\cite{rout2024rb} also fails to depict embroidery structures with an existing style descriptor, or preserve design content with the CLIP image encoder and attention feature aggregation. 
Moreover, we provide qualitative comparisons with Attention Distillation~\cite{zhou2025attention} and Analogist~\cite{gu2024analogist} in the supplementary material, showing that approaches relying on pretrained attention features or textural descriptions fail to capture fine-grained styles such as embroidery.

In Tab.~\ref{tab:combined_results},  Ours, DB-LoRA and B-LoRA have very similar results in HFRD and LPIPS, which is probably due to the limitations of current metrics in evaluating embroidery style and design content at a fine-grained level. 
InstantStyle achieves the best LPIPS as they mainly recover the input design. 
We attain the best score in Histogram Loss, demonstrating improved disentanglement from reference color. An ablation study on the color correction module and a discussion of metric limitations are included in the supplementary material. 
For text-based generation, our method achieves the highest CLIP-Score, indicating strong compliance with text prompts, and the highest Histogram Loss, reflecting minimal resemblance to the reference color.

Additionally, we compare to two methods for embroidery synthesis, both with three types of flat stitches. As in the last row of Fig.~\ref{fig:img2img1}, we can generate highly realistic embroideries tailored to different references. 
MSEmbGAN~\cite{hu2024msembgan} proposes a GAN-based approach and is bounded by the rendered training data. As their code and models are unavailable, we present a visual result generated with Wilcom EmbroideryStudio to approximate its upper bound, which exhibits a clear deficiency in photorealism. 
For the method of \cite{guan2021automatic}, we show their results in automatic mode (a) and long-stitch mode (b), which shows unnatural region division and rendering artifacts.

\subsection{Ablation Study}
\label{sec:ablation}
We conduct ablation studies to analyze the efficacy of the components of our method. Specifically, we compare three variations: (1) \textbf{2-Block Modulation}, where we use the two LoRA blocks with the lowest average cosine similarity to capture embroidery style instead of four; (2) \textbf{w/o Modulation}, where we use all LoRA blocks instead of four to capture the embroidery style; and (3) \textbf{w/o Contrastive Learning}, where we adopt results from our first-stage training. The comparisons are shown in Fig.~\ref{fig:analysis_ablation} (b). Using two blocks alone struggles to capture fine-grained embroidery structures, while using all blocks or omitting the second-stage contrastive learning fails to effectively decouple style from color and semantics, which causes unnatural fusion with the input design. We conduct user studies for quantitative evaluation due to the limitations of existing metrics, and provide additional ablation studies on using different blocks in the supplementary material.

\begin{table}
    \centering
    \caption{User studies on overall embroidery quality, style consistency, and design preservation. The numbers represent the percentage of votes that these methods are preferred over our final model. For DB-LoRA, 19.44\% in Quality means 19.44\% of votes favor DB-LoRA's embroidery quality over ours, while 80.56\% disagree.}
    \scalebox{0.9}{
    \begin{tabular}{lccccc}
        \toprule
        \textbf{Method} & \textbf{Quality (\%)} & \textbf{Style (\%)} & \textbf{Design (\%)} \\
        \midrule
        DB-LoRA            & 19.44 & 24.05 & 27.63 \\
        B-LoRA             & 11.59 & 16.00 & 14.29 \\
        InstantStyle       & 10.68 &  2.50 & 31.58 \\
        PairCustomization  &  2.67 &  1.25 & 13.89 \\
        StyleID            &  1.30 &  0.00 &  5.13 \\
        RB-Modulation      & 13.16 &  3.75 &  0.00 \\
        \midrule
        2-Block Modulation    & 10.81 &  7.59 &  8.97 \\
        w/o Modulation     &  9.46 &  7.79 &  7.79 \\
        w/o Contrastive Learning & 25.00 & 34.92 & 23.44 \\
        \bottomrule
    \end{tabular}}
\label{tab:user_study}
\end{table}

\begin{figure*}[t]
    \centering
    \includegraphics[width=\textwidth]{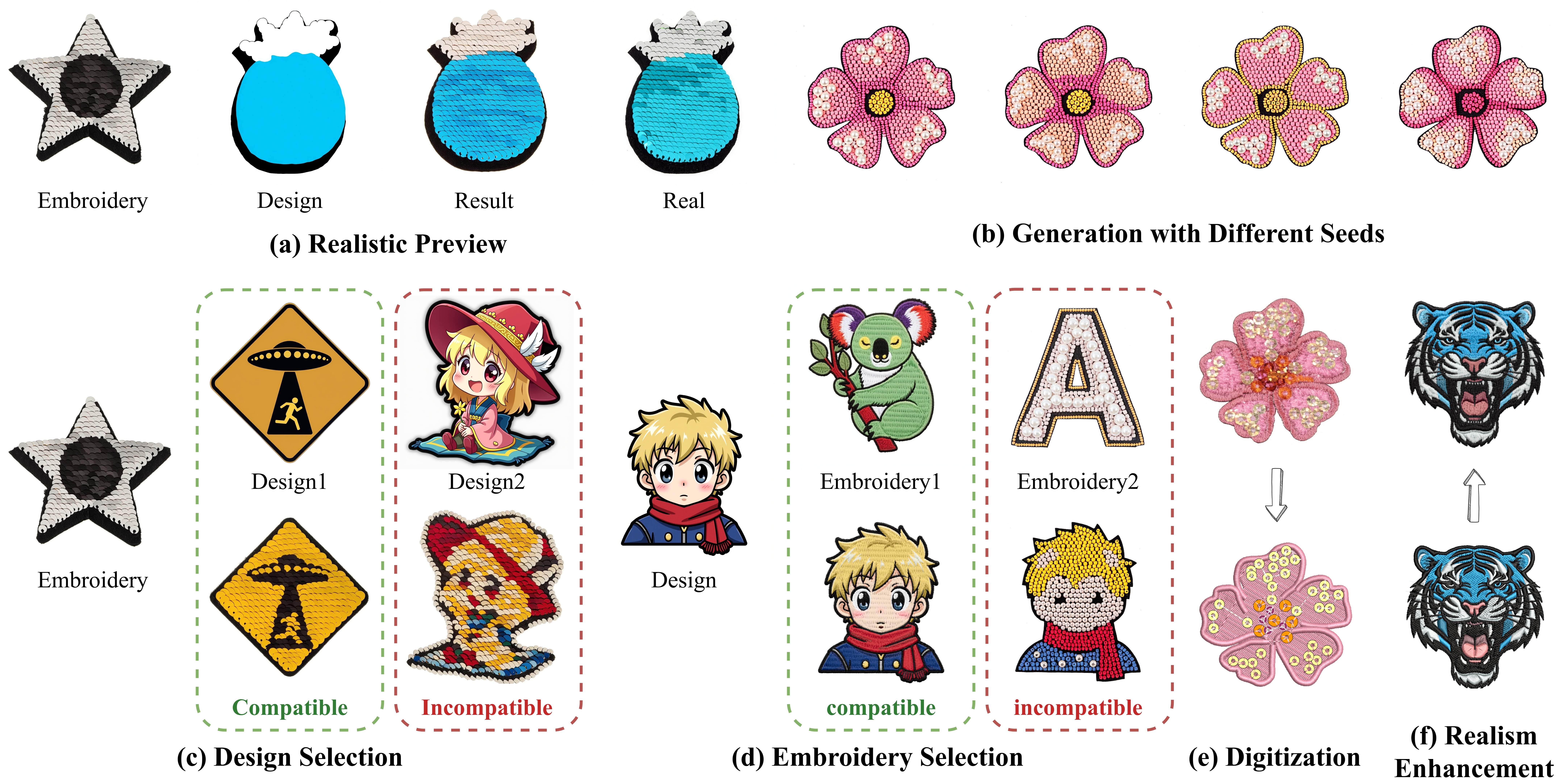}
    \caption{Applications of embroidery customization. (a) The result from our generation is as realistic as real embroidery. (b) Our method can generate diverse results using different random seeds, offering more options for production. (c) Given a reference embroidery, our method can help identify compatible designs. (d) Given a design image, our method can suggest compatible embroidery styles. (e) Digitization by tracing a generated embroidery with Wilcom EmbroideryStudio. (f) Our method can enhance the realism of Wilcom EmbroideryStudio renderings.}
    \label{fig:application}
\end{figure*}

\subsection{User Study}
\label{sec:user_study}
To compensate for the misalignment of existing metrics and real objectives for embroidery customization, we conduct user studies to compare our methods to previous works and ablation variations. Specifically, we follow similar setting as \cite{wang2023stylediffusion, jones2024customizing} and compare our method to others in pairs. 
For each user, we randomly sample 90 pairs, with each pair comprising a generated image from our method and one of another method. 
We provide the user study interface in the supplementary material. For each pair, users are asked to select their preferred option in terms of overall embroidery quality, style consistency with the reference, and design content preservation. Each question offers three choices: Method A, Method B, or Abstain.

We conducted an online questionnaire with 20 users, including two professionals in embroidery and 18 ordinary customers. Before answering, they viewed 10 reference embroidery images, followed by 10 generations with poor quality, inconsistent styles, or misaligned designs. 
All users completed the task within 30 minutes, though no time limit was set. 
The statistical results are presented in Tab.~\ref{tab:user_study}. Our method receives a clear preference over previous works, as well as over the ablation variations. Comparisons with other methods are based on valid votes. For Quality, Style, and Design, the valid ratios are 91.1\%, 96.2\%, and 94.2\%, respectively, with each category comprising 1,800 votes in total.

\subsection{Transformation to Embroidery Workflows}
\label{sec:application}
In this section, we illustrate the potential of our embroidery customization technique for transforming real-world embroidery workflows. Specifically, we first demonstrate its utility in enabling preview and presale, thereby bridging visual communication between producers and customers. We then showcase its role in fabrication support through embroidery digitization and visualization enhancement. Finally, we present more usage scenarios that highlight its capability to generate high-quality embroidery and design images.

\paragraph{Preview and Presale}
With our generated embroidery previews, producer and consumer preferences can be better aligned, facilitating more effective presale decisions. 
As shown in Fig.~\ref{fig:application} (a), we first verify that our generated results achieve realism comparable to real embroidery. Based on this, our method can then suggest compatible design patterns or suitable embroidery styles given a reference embroidery or a design image, as illustrated in Figs.~\ref{fig:application} (c) and (d). 
In the supplementary material, we conduct user studies to evaluate whether participants can distinguish between real and generated embroidery images, and quantitatively assess how previews using our generated results influence their preferences.

\begin{figure*}[t]
    \centering
    \includegraphics[width=\textwidth]{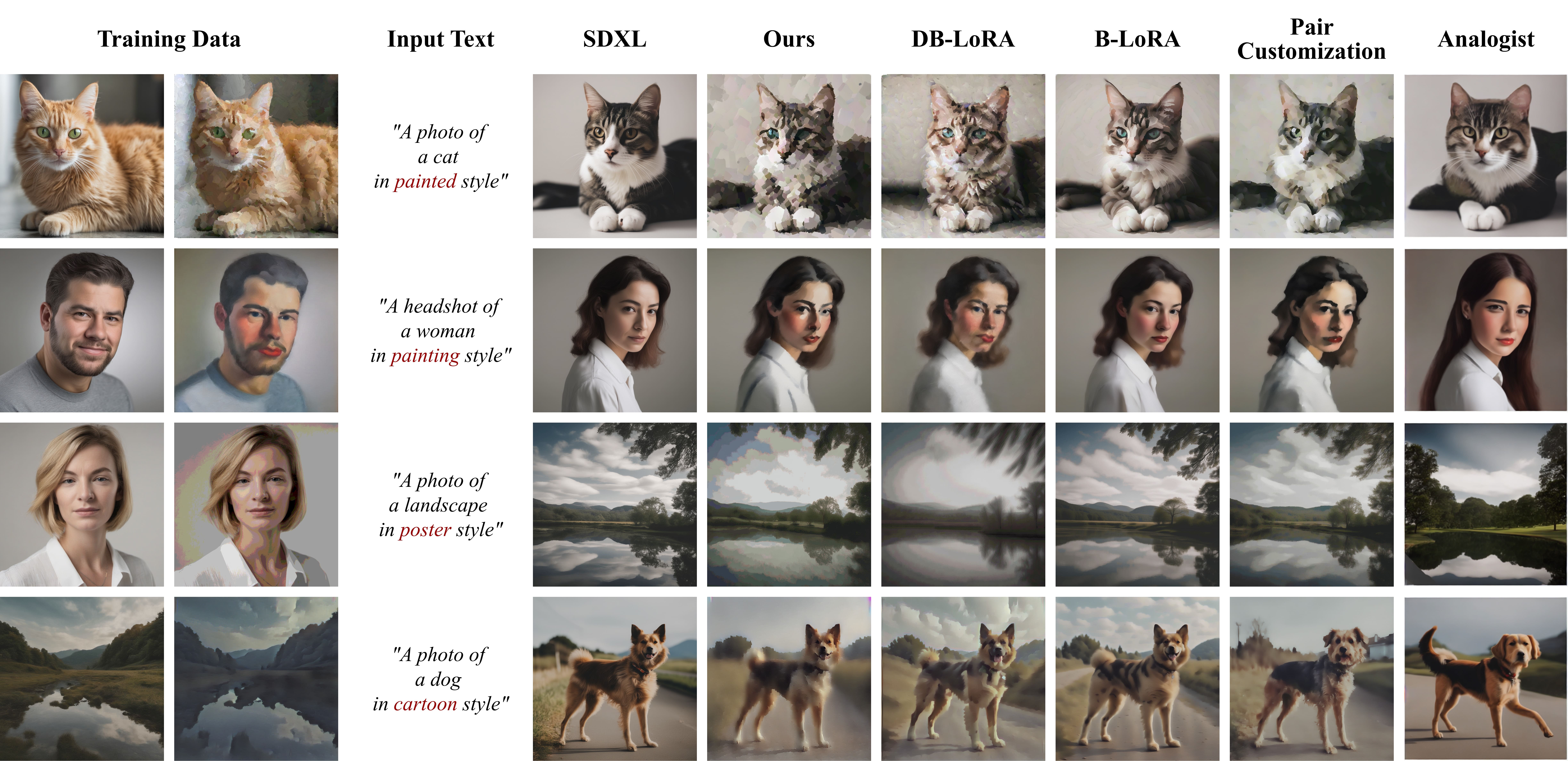}
    \caption{Comparison on four examples of artistic style transfer. The training data (Row 2) \textcopyright~ PairCustomization~\cite{jones2024customizing}.}
    \label{fig:artwork}
\end{figure*}

\paragraph{Fabrication Support}
Our embroidery customization technique supports the fabrication process in both digitization and visualization. Given a designated reference embroidery style and design image, our method can generate diverse outcomes (Fig.~\ref{fig:application} (b)), thereby reducing iterative cycles of digitization and confirmation. 
Once a design is finalized, it can be digitized using Wilcom EmbroideryStudio to obtain a manufacturable file (Fig.~\ref{fig:application} (e)). The tracing process involves color extraction, physical size alignment, layer-by-layer analysis, stitch filling, and decorative embellishment placement. Further details are provided in the supplementary document and video. This example was completed within 10 minutes, although additional manual refinement is required for the final production file. 
In addition, our customization can serve as a realistic rendering module (Fig.~\ref{fig:application} (f)), enhancing the realism of digitized embroidery and facilitating communication between producers and customers.

\paragraph{More Applications}
Our generated embroidery can be overlaid onto garments, bags, or hats to provide more intuitive visual previews, through integration with ACE++~\cite{mao2025ace++}, as shown in Fig.~\ref{fig:teaser}. Moreover, the capability of our method in generating high-quality embroidery data helps address the challenge of data scarcity in this domain. Additionally, our embroidery-to-design module effectively recovers well-aligned design images from reference embroideries, enabling novel style synthesis with consistent designs. Additional visual results and implementation details are provided in the supplementary material.

\subsection{Generalization to Other Styles}
\label{sec:other_style}
We evaluate our method across diverse styles to demonstrate its effectiveness in disentangling style and content. To this end, we compare against prior work on three tasks: artistic style transfer, sketch colorization, and appearance transfer. For each task, we construct a domain-specific data pair, followed by our standard block selection and contrastive LoRA learning. More results and implementation details are included in the supplementary material.

\paragraph{Photo to Artwork}
We compare to PairCustomization, which also learns artistic style from a single image pair. Following their setup, we construct photo–artwork pairs using external stylization methods (Fig.~\ref{fig:artwork}). The objective is to generate stylized images while preserving SDXL-generated content from text inputs, thereby verifying style–content disentanglement during learning. 
We compare with LoRA-based methods using a timestep-controlled LoRA activation strategy during denoising, following PairCustomization. As in Fig.~\ref{fig:artwork}, DB-LoRA suffers from artifacts due to overfitting to training content, while B-LoRA yields weak stylization. Our method is comparable to PairCustomization on in-domain cases (Rows 1–2) and performs slightly better on cross-domain cases (Rows 3–4). Additional qualitative and quantitative analysis are provided in the supplementary material. These results confirm that our method achieves effective style–content disentanglement and offers a viable alternative for this task. 
We also evaluate Analogist with SDXL generation as its input for image analogy, and the results highlight its limitations in transferring such fine-grained styles using textual descriptions.

\begin{figure*}
    \centering
    \includegraphics[width=\textwidth]{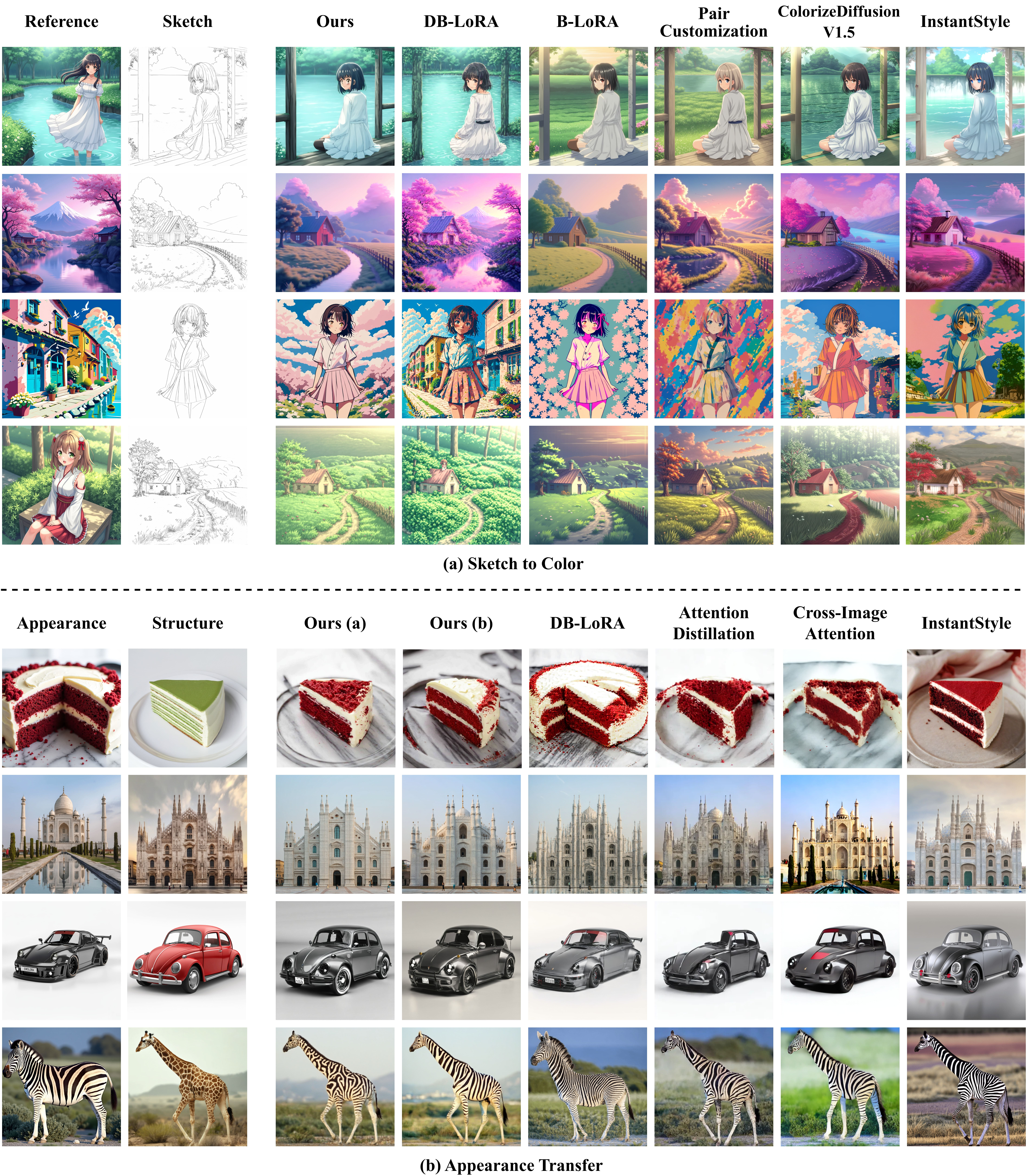}
    \caption{Sketch colorization and appearance transfer. (a) Comparison on four examples of sketch-to-color style transfer. (b) Comparison on four examples of appearance transfer. For zebra, the content Canny map contains stripe patterns, leading to a decoupling between appearance and structure. Ours (a) uses only style blocks and captures distorted strips, while Ours (b) leverages all blocks and restores complete stripe structure.}
    \label{fig:color_appearance}
\end{figure*}

\paragraph{Sketch to Color}
Our method can be applied to sketch colorization, through training on a color-Canny image pair to separate style (color and shading) from content (semantics and layout). 
In Fig.~\ref{fig:color_appearance} (a), Ours achieves color consistency with the reference and effective content compliance. 
In contrast, DB-LoRA shows content entanglement (e.g., generating "standing in water" instead of "sitting on the floor" in the first row), while B-LoRA, InstantStyle, and PairCustomization exhibits noticeable color drift from the reference. 
ColorizeDiffusion~\cite{yan2025image} is trained on millions of colorization samples.

\paragraph{Appearance Transfer}
We extend our method to appearance transfer by training on appearance–Canny pairs, where the style involves richer textures beyond color. 
At inference, we use HED~\cite{xie2015holistically} maps of structure images and apply either style blocks as Ours (a), or all blocks as Ours (b), as shown in Fig.~\ref{fig:color_appearance} (b). 
Ours (a) achieves better structural consistency through stronger appearance-structure disentanglement, while Ours (b) preserves partial reference structure and fuses it compatibly with the input.

\section{Limitation and Discussion}
\label{sec:limitation}
In this section, we present two representative failure cases, discuss the limitations of our method, and outline several future directions for supporting fabrication.

\begin{figure}
    \centering
    \includegraphics[width=0.48\textwidth]{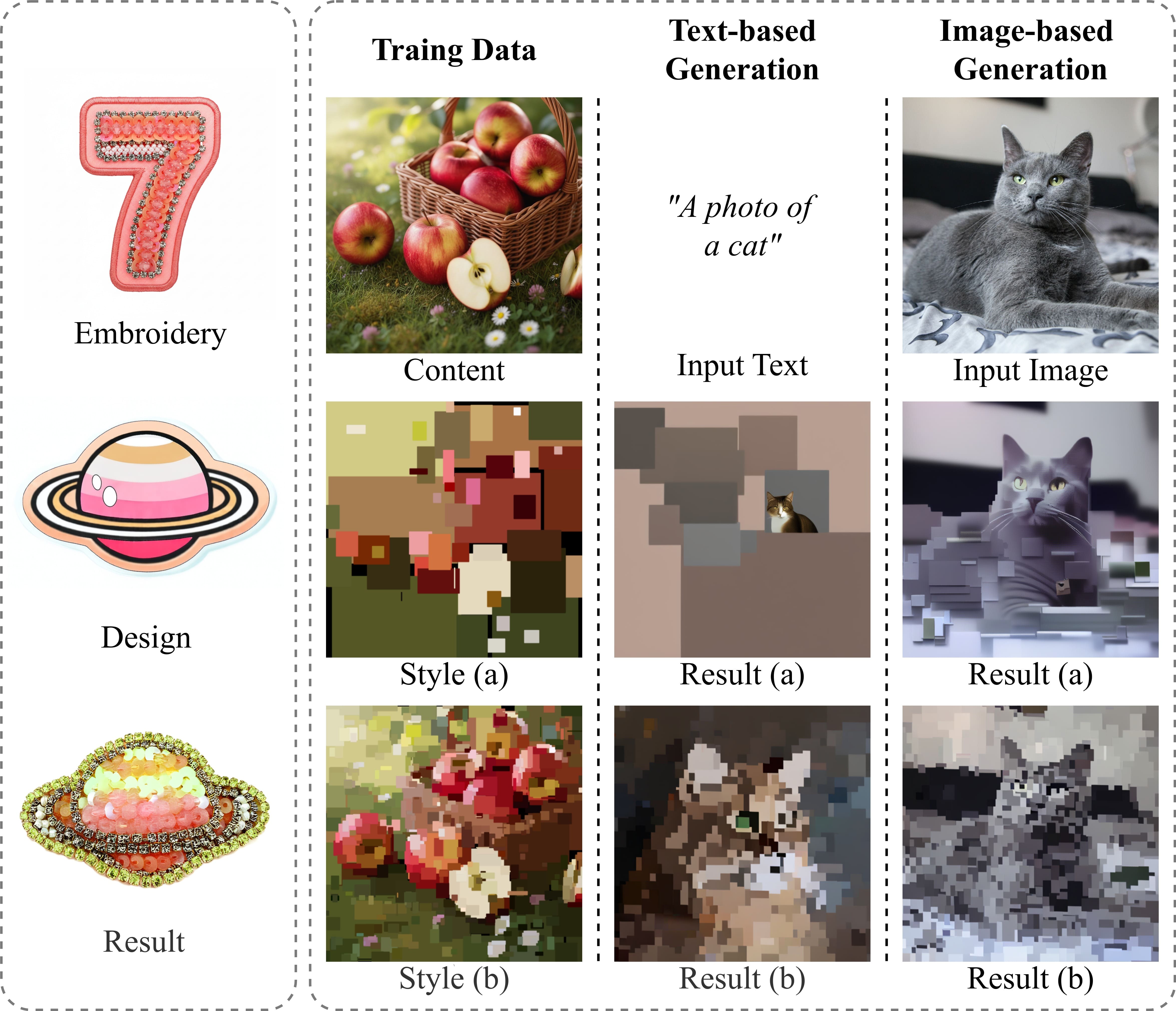}
    \caption{Two failure cases of our method when applied to highly complex or abstract styles. The input image (Row 1, Column 4) \textcopyright~
    \href{https://unsplash.com/photos/russian-blue-cat-on-white-and-blue-textile-JLN0U5Rj0gg}{Milada Vigerova}.}
    \label{fig:failure_case}
\end{figure}

\paragraph{Failure Cases}
Our method faces challenges with highly complex styles that combine multiple materials or overly abstract styles, as illustrated in Fig.~\ref{fig:failure_case}. 
For \textbf{embroidery customization}, intricate reference styles—such as intersecting combinations of multiple bead types—often yield impractical customization results, while the reflective characteristics of rhinestones introduce additional imaging difficulties. Furthermore, localized style specification or editing is not yet supported. 
For \textbf{general style}, we employ the rectangle renderer from Stylized Neural Painting \cite{zou2021stylized} to generate 8-bit artworks with two stroke configurations: (a) 50 strokes and (b) 550 strokes, as in Fig.~\ref{fig:failure_case} Style (a) and Style (b). 
Our method performs well when paired data are reasonably aligned, as in Style (b), but struggles with overly abstract styles such as Style (a), where block selection may fail to identify appropriate layers for modulation, leading to weak stylization and poor content blending.

\paragraph{Method Limitations}
Our method comprises multiple stages: pairwise data construction, network block selection, and two-stage training. 
We leverage SDXL for style–content disentanglement and SD3 as a plug-and-play module for design emulation, while the framework can be extended or unified with more powerful models exhibiting similar properties. 
We empirically select four blocks (e.g., 2, 3, 7, 8) to balance style completeness and style–content separation, though omitting blocks 1 and 11 can reduce color and appearance integrity in sketch colorization and appearance transfer. Our method could be enhanced with automatic block selection based on low cosine similarity and statistical constraints, and further refined to operate at a finer granularity through soft weighting rather than hard selection. 
In future work, we aim to reformulate fine-grained style customization within a meta-learning framework—e.g., treating the first training stage as meta-training to learn generalizable style disentanglement. 
Finally, when no reference embroidery is available, EmoLoRAs pretrained on tagged embroidery styles can serve as selectable modules, with text-only inputs mapped to style tags via LLMs, while combining style primitives from multiple references (e.g., chenille and sequin) remains an open direction.

\paragraph{Future Fabrication Support}
To support future fabrication, we outline several directions for automatically generating production-ready files with rich structured information. A central step is defining a representation for primitive embroidery instructions (EmbIns)—including stitch coordinates, needle commands, and color sequences—analogous to SVG but with greater complexity. 
One promising avenue is the development of a differentiable rasterizer for EmbIns, akin to DiffVG~\cite{li2020differentiable}, enabling optimization-based generation of EmbIns from customized images. This approach could be extended to a unified framework for joint image and EmbIns generation, similar to VectorFusion~\cite{jain2023vectorfusion}. 
Another direction is a multi-modal approach that tokenizes EmbIns for joint learning with text and images, as in OmniSVG~\cite{yang2025omnisvg}, though this would require large-scale datasets for effective training.

\section{Conclusion}
\label{sec:conclusion}
In this paper, we address fine-grained style customization by introducing a contrastive learning framework that disentangles style and content from a single reference image, based on the classic concept of image analogy and leveraging decoupled representations from pretrained diffusion models. To capture fine-grained style features, we propose a two-stage contrastive LoRA modulation technique, EmoLoRA, which mitigates data scarcity through self-knowledge distillation. Our approach significantly outperforms existing methods in embroidery customization, with extensive analysis of its potential to transform real-world embroidery workflows. Moreover, it generalizes to three additional visual attribute transfer tasks, providing a new alternative to existing works.

For future work, we envision unifying pair-wise data generation with style customization into a single framework, enhancing block selection through automatic or soft-weighted strategies, and reformulating the two-stage learning as a meta-learning framework. Additionally, we outline directions for automatically generating production-ready embroidery files, drawing inspiration from SVG representations, including defining primitive embroidery instructions, developing differentiable rasterizers, and exploring joint learning of images, text, and embroidery instructions.

\begin{acks}
This work was supported by Key R\&D Program of Zhejiang (No. 2023C01047) and the Ningbo Major Special Projects of the "Science and Technology Innovation 2025" (Grant No. 2023Z143).
\end{acks}

\bibliographystyle{ACM-Reference-Format}
\bibliography{sample-bibliography}


\begin{thebibliography}{100}


\ifx \showCODEN    \undefined \def \showCODEN     #1{\unskip}     \fi
\ifx \showDOI      \undefined \def \showDOI       #1{#1}\fi
\ifx \showISBNx    \undefined \def \showISBNx     #1{\unskip}     \fi
\ifx \showISBNxiii \undefined \def \showISBNxiii  #1{\unskip}     \fi
\ifx \showISSN     \undefined \def \showISSN      #1{\unskip}     \fi
\ifx \showLCCN     \undefined \def \showLCCN      #1{\unskip}     \fi
\ifx \shownote     \undefined \def \shownote      #1{#1}          \fi
\ifx \showarticletitle \undefined \def \showarticletitle #1{#1}   \fi
\ifx \showURL      \undefined \def \showURL       {\relax}        \fi
\providecommand\bibfield[2]{#2}
\providecommand\bibinfo[2]{#2}
\providecommand\natexlab[1]{#1}
\providecommand\showeprint[2][]{arXiv:#2}

\bibitem[Achiam et~al\mbox{.}(2023)]%
        {achiam2023gpt}
\bibfield{author}{\bibinfo{person}{Josh Achiam}, \bibinfo{person}{Steven Adler}, \bibinfo{person}{Sandhini Agarwal}, \bibinfo{person}{Lama Ahmad}, \bibinfo{person}{Ilge Akkaya}, \bibinfo{person}{Florencia~Leoni Aleman}, \bibinfo{person}{Diogo Almeida}, \bibinfo{person}{Janko Altenschmidt}, \bibinfo{person}{Sam Altman}, \bibinfo{person}{Shyamal Anadkat}, {et~al\mbox{.}}} \bibinfo{year}{2023}\natexlab{}.
\newblock \showarticletitle{Gpt-4 technical report}.
\newblock \bibinfo{journal}{\emph{arXiv preprint arXiv:2303.08774}} (\bibinfo{year}{2023}).
\newblock


\bibitem[Afifi et~al\mbox{.}(2021)]%
        {afifi2021histogan}
\bibfield{author}{\bibinfo{person}{Mahmoud Afifi}, \bibinfo{person}{Marcus~A Brubaker}, {and} \bibinfo{person}{Michael~S Brown}.} \bibinfo{year}{2021}\natexlab{}.
\newblock \showarticletitle{Histogan: Controlling colors of gan-generated and real images via color histograms}. In \bibinfo{booktitle}{\emph{Proceedings of the IEEE/CVF Conference on Computer Vision and Pattern Recognition}}. \bibinfo{pages}{7941--7950}.
\newblock


\bibitem[Alaluf et~al\mbox{.}(2024)]%
        {alaluf2024cross}
\bibfield{author}{\bibinfo{person}{Yuval Alaluf}, \bibinfo{person}{Daniel Garibi}, \bibinfo{person}{Or Patashnik}, \bibinfo{person}{Hadar Averbuch-Elor}, {and} \bibinfo{person}{Daniel Cohen-Or}.} \bibinfo{year}{2024}\natexlab{}.
\newblock \showarticletitle{Cross-image attention for zero-shot appearance transfer}. In \bibinfo{booktitle}{\emph{ACM SIGGRAPH 2024 Conference Papers}}. \bibinfo{pages}{1--12}.
\newblock


\bibitem[Brooks et~al\mbox{.}(2023)]%
        {brooks2023instructpix2pix}
\bibfield{author}{\bibinfo{person}{Tim Brooks}, \bibinfo{person}{Aleksander Holynski}, {and} \bibinfo{person}{Alexei~A Efros}.} \bibinfo{year}{2023}\natexlab{}.
\newblock \showarticletitle{Instructpix2pix: Learning to follow image editing instructions}. In \bibinfo{booktitle}{\emph{Proceedings of the IEEE/CVF Conference on Computer Vision and Pattern Recognition}}. \bibinfo{pages}{18392--18402}.
\newblock


\bibitem[Chen et~al\mbox{.}(2012)]%
        {chen2012embroidery}
\bibfield{author}{\bibinfo{person}{Xinling Chen}, \bibinfo{person}{Michael McCool}, \bibinfo{person}{Asanobu Kitamoto}, {and} \bibinfo{person}{Stephen Mann}.} \bibinfo{year}{2012}\natexlab{}.
\newblock \showarticletitle{Embroidery modeling and rendering}.
\newblock In \bibinfo{booktitle}{\emph{Proceedings of Graphics Interface 2012}}. \bibinfo{pages}{131--139}.
\newblock


\bibitem[Chung et~al\mbox{.}(2024)]%
        {chung2024style}
\bibfield{author}{\bibinfo{person}{Jiwoo Chung}, \bibinfo{person}{Sangeek Hyun}, {and} \bibinfo{person}{Jae-Pil Heo}.} \bibinfo{year}{2024}\natexlab{}.
\newblock \showarticletitle{Style injection in diffusion: A training-free approach for adapting large-scale diffusion models for style transfer}. In \bibinfo{booktitle}{\emph{Proceedings of the IEEE/CVF Conference on Computer Vision and Pattern Recognition}}. \bibinfo{pages}{8795--8805}.
\newblock


\bibitem[Cui et~al\mbox{.}(2017)]%
        {cui2017image}
\bibfield{author}{\bibinfo{person}{Dele Cui}, \bibinfo{person}{Yun Sheng}, {and} \bibinfo{person}{Guixu Zhang}.} \bibinfo{year}{2017}\natexlab{}.
\newblock \showarticletitle{Image-based embroidery modeling and rendering}.
\newblock \bibinfo{journal}{\emph{Computer Animation and Virtual Worlds}} \bibinfo{volume}{28}, \bibinfo{number}{2} (\bibinfo{year}{2017}), \bibinfo{pages}{e1725}.
\newblock


\bibitem[Dalva and Yanardag(2024)]%
        {dalva2024noiseclr}
\bibfield{author}{\bibinfo{person}{Yusuf Dalva} {and} \bibinfo{person}{Pinar Yanardag}.} \bibinfo{year}{2024}\natexlab{}.
\newblock \showarticletitle{Noiseclr: A contrastive learning approach for unsupervised discovery of interpretable directions in diffusion models}. In \bibinfo{booktitle}{\emph{Proceedings of the IEEE/CVF Conference on Computer Vision and Pattern Recognition}}. \bibinfo{pages}{24209--24218}.
\newblock


\bibitem[Deng et~al\mbox{.}(2023)]%
        {deng2023z}
\bibfield{author}{\bibinfo{person}{Yingying Deng}, \bibinfo{person}{Xiangyu He}, \bibinfo{person}{Fan Tang}, {and} \bibinfo{person}{Weiming Dong}.} \bibinfo{year}{2023}\natexlab{}.
\newblock \showarticletitle{Z*: Zero-shot Style Transfer via Attention Rearrangement}.
\newblock \bibinfo{journal}{\emph{arXiv preprint arXiv:2311.16491}} (\bibinfo{year}{2023}).
\newblock


\bibitem[Deng et~al\mbox{.}(2022)]%
        {deng2022stytr2}
\bibfield{author}{\bibinfo{person}{Yingying Deng}, \bibinfo{person}{Fan Tang}, \bibinfo{person}{Weiming Dong}, \bibinfo{person}{Chongyang Ma}, \bibinfo{person}{Xingjia Pan}, \bibinfo{person}{Lei Wang}, {and} \bibinfo{person}{Changsheng Xu}.} \bibinfo{year}{2022}\natexlab{}.
\newblock \showarticletitle{Stytr2: Image style transfer with transformers}. In \bibinfo{booktitle}{\emph{Proceedings of the IEEE/CVF Conference on Computer Vision and Pattern Recognition}}. \bibinfo{pages}{11326--11336}.
\newblock


\bibitem[Dhariwal and Nichol(2021)]%
        {dhariwal2021diffusion}
\bibfield{author}{\bibinfo{person}{Prafulla Dhariwal} {and} \bibinfo{person}{Alexander Nichol}.} \bibinfo{year}{2021}\natexlab{}.
\newblock \showarticletitle{Diffusion models beat gans on image synthesis}.
\newblock \bibinfo{journal}{\emph{Advances in Neural Information Processing Systems}}  \bibinfo{volume}{34} (\bibinfo{year}{2021}), \bibinfo{pages}{8780--8794}.
\newblock


\bibitem[Efros and Freeman(2001)]%
        {efros2001image}
\bibfield{author}{\bibinfo{person}{Alexei~A Efros} {and} \bibinfo{person}{William~T Freeman}.} \bibinfo{year}{2001}\natexlab{}.
\newblock \showarticletitle{Image quilting for texture synthesis and transfer}. In \bibinfo{booktitle}{\emph{Proceedings of the 28th Annual Conference on Computer Graphics and Interactive Techniques}}. \bibinfo{pages}{341--346}.
\newblock


\bibitem[Esser et~al\mbox{.}(2024)]%
        {esser2024scaling}
\bibfield{author}{\bibinfo{person}{Patrick Esser}, \bibinfo{person}{Sumith Kulal}, \bibinfo{person}{Andreas Blattmann}, \bibinfo{person}{Rahim Entezari}, \bibinfo{person}{Jonas M{\"u}ller}, \bibinfo{person}{Harry Saini}, \bibinfo{person}{Yam Levi}, \bibinfo{person}{Dominik Lorenz}, \bibinfo{person}{Axel Sauer}, \bibinfo{person}{Frederic Boesel}, {et~al\mbox{.}}} \bibinfo{year}{2024}\natexlab{}.
\newblock \showarticletitle{Scaling rectified flow transformers for high-resolution image synthesis}. In \bibinfo{booktitle}{\emph{Forty-first International Conference on Machine Learning}}.
\newblock


\bibitem[Esser et~al\mbox{.}(2021)]%
        {esser2021taming}
\bibfield{author}{\bibinfo{person}{Patrick Esser}, \bibinfo{person}{Robin Rombach}, {and} \bibinfo{person}{Bjorn Ommer}.} \bibinfo{year}{2021}\natexlab{}.
\newblock \showarticletitle{Taming transformers for high-resolution image synthesis}. In \bibinfo{booktitle}{\emph{Proceedings of the IEEE/CVF Conference on Computer Vision and Pattern Recognition}}. \bibinfo{pages}{12873--12883}.
\newblock


\bibitem[Frenkel et~al\mbox{.}(2025)]%
        {frenkel2025implicit}
\bibfield{author}{\bibinfo{person}{Yarden Frenkel}, \bibinfo{person}{Yael Vinker}, \bibinfo{person}{Ariel Shamir}, {and} \bibinfo{person}{Daniel Cohen-Or}.} \bibinfo{year}{2025}\natexlab{}.
\newblock \showarticletitle{Implicit style-content separation using b-lora}. In \bibinfo{booktitle}{\emph{European Conference on Computer Vision}}. Springer, \bibinfo{pages}{181--198}.
\newblock


\bibitem[Gal et~al\mbox{.}(2022)]%
        {gal2022image}
\bibfield{author}{\bibinfo{person}{Rinon Gal}, \bibinfo{person}{Yuval Alaluf}, \bibinfo{person}{Yuval Atzmon}, \bibinfo{person}{Or Patashnik}, \bibinfo{person}{Amit~H Bermano}, \bibinfo{person}{Gal Chechik}, {and} \bibinfo{person}{Daniel Cohen-Or}.} \bibinfo{year}{2022}\natexlab{}.
\newblock \showarticletitle{An image is worth one word: Personalizing text-to-image generation using textual inversion}.
\newblock \bibinfo{journal}{\emph{arXiv preprint arXiv:2208.01618}} (\bibinfo{year}{2022}).
\newblock


\bibitem[Garibi et~al\mbox{.}(2024)]%
        {garibi2024renoise}
\bibfield{author}{\bibinfo{person}{Daniel Garibi}, \bibinfo{person}{Or Patashnik}, \bibinfo{person}{Andrey Voynov}, \bibinfo{person}{Hadar Averbuch-Elor}, {and} \bibinfo{person}{Daniel Cohen-Or}.} \bibinfo{year}{2024}\natexlab{}.
\newblock \showarticletitle{ReNoise: Real Image Inversion Through Iterative Noising}.
\newblock \bibinfo{journal}{\emph{arXiv preprint arXiv:2403.14602}} (\bibinfo{year}{2024}).
\newblock


\bibitem[Gatys et~al\mbox{.}(2016)]%
        {gatys2016image}
\bibfield{author}{\bibinfo{person}{Leon~A Gatys}, \bibinfo{person}{Alexander~S Ecker}, {and} \bibinfo{person}{Matthias Bethge}.} \bibinfo{year}{2016}\natexlab{}.
\newblock \showarticletitle{Image style transfer using convolutional neural networks}. In \bibinfo{booktitle}{\emph{Proceedings of the IEEE Conference on Computer Vision and Pattern Recognition}}. \bibinfo{pages}{2414--2423}.
\newblock


\bibitem[Goodfellow et~al\mbox{.}(2014)]%
        {goodfellow2014generative}
\bibfield{author}{\bibinfo{person}{Ian Goodfellow}, \bibinfo{person}{Jean Pouget-Abadie}, \bibinfo{person}{Mehdi Mirza}, \bibinfo{person}{Bing Xu}, \bibinfo{person}{David Warde-Farley}, \bibinfo{person}{Sherjil Ozair}, \bibinfo{person}{Aaron Courville}, {and} \bibinfo{person}{Yoshua Bengio}.} \bibinfo{year}{2014}\natexlab{}.
\newblock \showarticletitle{Generative adversarial nets}.
\newblock \bibinfo{journal}{\emph{Advances in Neural Information Processing Systems}}  \bibinfo{volume}{27} (\bibinfo{year}{2014}).
\newblock


\bibitem[Gu et~al\mbox{.}(2024)]%
        {gu2024analogist}
\bibfield{author}{\bibinfo{person}{Zheng Gu}, \bibinfo{person}{Shiyuan Yang}, \bibinfo{person}{Jing Liao}, \bibinfo{person}{Jing Huo}, {and} \bibinfo{person}{Yang Gao}.} \bibinfo{year}{2024}\natexlab{}.
\newblock \showarticletitle{Analogist: Out-of-the-box visual in-context learning with image diffusion model}.
\newblock \bibinfo{journal}{\emph{ACM Transactions on Graphics (TOG)}} \bibinfo{volume}{43}, \bibinfo{number}{4} (\bibinfo{year}{2024}), \bibinfo{pages}{1--15}.
\newblock


\bibitem[Guan et~al\mbox{.}(2021)]%
        {guan2021automatic}
\bibfield{author}{\bibinfo{person}{Xinyang Guan}, \bibinfo{person}{Likang Luo}, \bibinfo{person}{Honglin Li}, \bibinfo{person}{He Wang}, \bibinfo{person}{Chen Liu}, \bibinfo{person}{Su Wang}, {and} \bibinfo{person}{Xiaogang Jin}.} \bibinfo{year}{2021}\natexlab{}.
\newblock \showarticletitle{Automatic embroidery texture synthesis for garment design and online display}.
\newblock \bibinfo{journal}{\emph{The Visual Computer}} \bibinfo{volume}{37}, \bibinfo{number}{9} (\bibinfo{year}{2021}), \bibinfo{pages}{2553--2565}.
\newblock


\bibitem[Hertz et~al\mbox{.}(2022)]%
        {hertz2022prompt}
\bibfield{author}{\bibinfo{person}{Amir Hertz}, \bibinfo{person}{Ron Mokady}, \bibinfo{person}{Jay Tenenbaum}, \bibinfo{person}{Kfir Aberman}, \bibinfo{person}{Yael Pritch}, {and} \bibinfo{person}{Daniel Cohen-Or}.} \bibinfo{year}{2022}\natexlab{}.
\newblock \showarticletitle{Prompt-to-prompt image editing with cross attention control}.
\newblock \bibinfo{journal}{\emph{arXiv preprint arXiv:2208.01626}} (\bibinfo{year}{2022}).
\newblock


\bibitem[Hertz et~al\mbox{.}(2024)]%
        {hertz2024style}
\bibfield{author}{\bibinfo{person}{Amir Hertz}, \bibinfo{person}{Andrey Voynov}, \bibinfo{person}{Shlomi Fruchter}, {and} \bibinfo{person}{Daniel Cohen-Or}.} \bibinfo{year}{2024}\natexlab{}.
\newblock \showarticletitle{Style aligned image generation via shared attention}. In \bibinfo{booktitle}{\emph{Proceedings of the IEEE/CVF Conference on Computer Vision and Pattern Recognition}}. \bibinfo{pages}{4775--4785}.
\newblock


\bibitem[Hertzmann et~al\mbox{.}(2001)]%
        {hertzmann2001image}
\bibfield{author}{\bibinfo{person}{Aaron Hertzmann}, \bibinfo{person}{Charles~E Jacobs}, \bibinfo{person}{Nuria Oliver}, \bibinfo{person}{Brian Curless}, {and} \bibinfo{person}{David~H Salesin}.} \bibinfo{year}{2001}\natexlab{}.
\newblock \showarticletitle{Image analogies}. In \bibinfo{booktitle}{\emph{Proceedings of the 28th Annual Conference on Computer Graphics and Interactive Techniques}}. \bibinfo{pages}{327--340}.
\newblock


\bibitem[Ho et~al\mbox{.}(2020)]%
        {ho2020denoising}
\bibfield{author}{\bibinfo{person}{Jonathan Ho}, \bibinfo{person}{Ajay Jain}, {and} \bibinfo{person}{Pieter Abbeel}.} \bibinfo{year}{2020}\natexlab{}.
\newblock \showarticletitle{Denoising diffusion probabilistic models}.
\newblock \bibinfo{journal}{\emph{Advances in Neural Information Processing Systems}}  \bibinfo{volume}{33} (\bibinfo{year}{2020}), \bibinfo{pages}{6840--6851}.
\newblock


\bibitem[Houlsby et~al\mbox{.}(2019)]%
        {houlsby2019parameter}
\bibfield{author}{\bibinfo{person}{Neil Houlsby}, \bibinfo{person}{Andrei Giurgiu}, \bibinfo{person}{Stanislaw Jastrzebski}, \bibinfo{person}{Bruna Morrone}, \bibinfo{person}{Quentin De~Laroussilhe}, \bibinfo{person}{Andrea Gesmundo}, \bibinfo{person}{Mona Attariyan}, {and} \bibinfo{person}{Sylvain Gelly}.} \bibinfo{year}{2019}\natexlab{}.
\newblock \showarticletitle{Parameter-efficient transfer learning for NLP}. In \bibinfo{booktitle}{\emph{International Conference on Machine Learning}}. PMLR, \bibinfo{pages}{2790--2799}.
\newblock


\bibitem[Hu et~al\mbox{.}(2021)]%
        {hu2021lora}
\bibfield{author}{\bibinfo{person}{Edward~J Hu}, \bibinfo{person}{Yelong Shen}, \bibinfo{person}{Phillip Wallis}, \bibinfo{person}{Zeyuan Allen-Zhu}, \bibinfo{person}{Yuanzhi Li}, \bibinfo{person}{Shean Wang}, \bibinfo{person}{Lu Wang}, {and} \bibinfo{person}{Weizhu Chen}.} \bibinfo{year}{2021}\natexlab{}.
\newblock \showarticletitle{Lora: Low-rank adaptation of large language models}.
\newblock \bibinfo{journal}{\emph{arXiv preprint arXiv:2106.09685}} (\bibinfo{year}{2021}).
\newblock


\bibitem[Hu et~al\mbox{.}(2024)]%
        {hu2024msembgan}
\bibfield{author}{\bibinfo{person}{Xinrong Hu}, \bibinfo{person}{Chen Yang}, \bibinfo{person}{Fei Fang}, \bibinfo{person}{Jin Huang}, \bibinfo{person}{Ping Li}, \bibinfo{person}{Bin ShengB}, {and} \bibinfo{person}{Tong-Yee Lee}.} \bibinfo{year}{2024}\natexlab{}.
\newblock \showarticletitle{Msembgan: Multi-stitch embroidery synthesis via region-aware texture generation}.
\newblock \bibinfo{journal}{\emph{IEEE Transactions on Visualization and Computer Graphics}} (\bibinfo{year}{2024}).
\newblock


\bibitem[Huang and Belongie(2017)]%
        {huang2017arbitrary}
\bibfield{author}{\bibinfo{person}{Xun Huang} {and} \bibinfo{person}{Serge Belongie}.} \bibinfo{year}{2017}\natexlab{}.
\newblock \showarticletitle{Arbitrary style transfer in real-time with adaptive instance normalization}. In \bibinfo{booktitle}{\emph{Proceedings of the IEEE International Conference on Computer Vision}}. \bibinfo{pages}{1501--1510}.
\newblock


\bibitem[Isola et~al\mbox{.}(2017)]%
        {isola2017image}
\bibfield{author}{\bibinfo{person}{Phillip Isola}, \bibinfo{person}{Jun-Yan Zhu}, \bibinfo{person}{Tinghui Zhou}, {and} \bibinfo{person}{Alexei~A Efros}.} \bibinfo{year}{2017}\natexlab{}.
\newblock \showarticletitle{Image-to-image translation with conditional adversarial networks}. In \bibinfo{booktitle}{\emph{Proceedings of the IEEE Conference on Computer Vision and Pattern Recognition}}. \bibinfo{pages}{1125--1134}.
\newblock


\bibitem[Jain et~al\mbox{.}(2023)]%
        {jain2023vectorfusion}
\bibfield{author}{\bibinfo{person}{Ajay Jain}, \bibinfo{person}{Amber Xie}, {and} \bibinfo{person}{Pieter Abbeel}.} \bibinfo{year}{2023}\natexlab{}.
\newblock \showarticletitle{Vectorfusion: Text-to-svg by abstracting pixel-based diffusion models}. In \bibinfo{booktitle}{\emph{Proceedings of the IEEE/CVF Conference on Computer Vision and Pattern Recognition}}. \bibinfo{pages}{1911--1920}.
\newblock


\bibitem[Johnson et~al\mbox{.}(2016)]%
        {johnson2016perceptual}
\bibfield{author}{\bibinfo{person}{Justin Johnson}, \bibinfo{person}{Alexandre Alahi}, {and} \bibinfo{person}{Li Fei-Fei}.} \bibinfo{year}{2016}\natexlab{}.
\newblock \showarticletitle{Perceptual losses for real-time style transfer and super-resolution}. In \bibinfo{booktitle}{\emph{Computer Vision--ECCV 2016: 14th European Conference, Amsterdam, The Netherlands, October 11-14, 2016, Proceedings, Part II 14}}. Springer, \bibinfo{pages}{694--711}.
\newblock


\bibitem[Jones et~al\mbox{.}(2024)]%
        {jones2024customizing}
\bibfield{author}{\bibinfo{person}{Maxwell Jones}, \bibinfo{person}{Sheng-Yu Wang}, \bibinfo{person}{Nupur Kumari}, \bibinfo{person}{David Bau}, {and} \bibinfo{person}{Jun-Yan Zhu}.} \bibinfo{year}{2024}\natexlab{}.
\newblock \showarticletitle{Customizing text-to-image models with a single image pair}. In \bibinfo{booktitle}{\emph{SIGGRAPH Asia 2024 Conference Papers}}. \bibinfo{pages}{1--13}.
\newblock


\bibitem[Karras(2019)]%
        {karras2019style}
\bibfield{author}{\bibinfo{person}{Tero Karras}.} \bibinfo{year}{2019}\natexlab{}.
\newblock \showarticletitle{A Style-Based Generator Architecture for Generative Adversarial Networks}.
\newblock \bibinfo{journal}{\emph{arXiv preprint arXiv:1812.04948}} (\bibinfo{year}{2019}).
\newblock


\bibitem[Karras et~al\mbox{.}(2020)]%
        {karras2020analyzing}
\bibfield{author}{\bibinfo{person}{Tero Karras}, \bibinfo{person}{Samuli Laine}, \bibinfo{person}{Miika Aittala}, \bibinfo{person}{Janne Hellsten}, \bibinfo{person}{Jaakko Lehtinen}, {and} \bibinfo{person}{Timo Aila}.} \bibinfo{year}{2020}\natexlab{}.
\newblock \showarticletitle{Analyzing and improving the image quality of stylegan}. In \bibinfo{booktitle}{\emph{Proceedings of the IEEE/CVF Conference on Computer Vision and Pattern Recognition}}. \bibinfo{pages}{8110--8119}.
\newblock


\bibitem[Kawar et~al\mbox{.}(2023)]%
        {kawar2023imagic}
\bibfield{author}{\bibinfo{person}{Bahjat Kawar}, \bibinfo{person}{Shiran Zada}, \bibinfo{person}{Oran Lang}, \bibinfo{person}{Omer Tov}, \bibinfo{person}{Huiwen Chang}, \bibinfo{person}{Tali Dekel}, \bibinfo{person}{Inbar Mosseri}, {and} \bibinfo{person}{Michal Irani}.} \bibinfo{year}{2023}\natexlab{}.
\newblock \showarticletitle{Imagic: Text-based real image editing with diffusion models}. In \bibinfo{booktitle}{\emph{Proceedings of the IEEE/CVF Conference on Computer Vision and Pattern Recognition}}. \bibinfo{pages}{6007--6017}.
\newblock


\bibitem[Kingma(2013)]%
        {kingma2013auto}
\bibfield{author}{\bibinfo{person}{Diederik~P Kingma}.} \bibinfo{year}{2013}\natexlab{}.
\newblock \showarticletitle{Auto-encoding variational bayes}.
\newblock \bibinfo{journal}{\emph{arXiv preprint arXiv:1312.6114}} (\bibinfo{year}{2013}).
\newblock


\bibitem[Kumari et~al\mbox{.}(2023)]%
        {kumari2023multi}
\bibfield{author}{\bibinfo{person}{Nupur Kumari}, \bibinfo{person}{Bingliang Zhang}, \bibinfo{person}{Richard Zhang}, \bibinfo{person}{Eli Shechtman}, {and} \bibinfo{person}{Jun-Yan Zhu}.} \bibinfo{year}{2023}\natexlab{}.
\newblock \showarticletitle{Multi-concept customization of text-to-image diffusion}. In \bibinfo{booktitle}{\emph{Proceedings of the IEEE/CVF Conference on Computer Vision and Pattern Recognition}}. \bibinfo{pages}{1931--1941}.
\newblock


\bibitem[Kwon and Ye(2022)]%
        {kwon2022diffusion}
\bibfield{author}{\bibinfo{person}{Gihyun Kwon} {and} \bibinfo{person}{Jong~Chul Ye}.} \bibinfo{year}{2022}\natexlab{}.
\newblock \showarticletitle{Diffusion-based image translation using disentangled style and content representation}.
\newblock \bibinfo{journal}{\emph{arXiv preprint arXiv:2209.15264}} (\bibinfo{year}{2022}).
\newblock


\bibitem[Li et~al\mbox{.}(2020)]%
        {li2020differentiable}
\bibfield{author}{\bibinfo{person}{Tzu-Mao Li}, \bibinfo{person}{Michal Luk{\'a}{\v{c}}}, \bibinfo{person}{Micha{\"e}l Gharbi}, {and} \bibinfo{person}{Jonathan Ragan-Kelley}.} \bibinfo{year}{2020}\natexlab{}.
\newblock \showarticletitle{Differentiable vector graphics rasterization for editing and learning}.
\newblock \bibinfo{journal}{\emph{ACM Transactions on Graphics (TOG)}} \bibinfo{volume}{39}, \bibinfo{number}{6} (\bibinfo{year}{2020}), \bibinfo{pages}{1--15}.
\newblock


\bibitem[Li et~al\mbox{.}(2025)]%
        {li2025styletokenizer}
\bibfield{author}{\bibinfo{person}{Wen Li}, \bibinfo{person}{Muyuan Fang}, \bibinfo{person}{Cheng Zou}, \bibinfo{person}{Biao Gong}, \bibinfo{person}{Ruobing Zheng}, \bibinfo{person}{Meng Wang}, \bibinfo{person}{Jingdong Chen}, {and} \bibinfo{person}{Ming Yang}.} \bibinfo{year}{2025}\natexlab{}.
\newblock \showarticletitle{StyleTokenizer: Defining Image Style by a Single Instance for Controlling Diffusion Models}. In \bibinfo{booktitle}{\emph{European Conference on Computer Vision}}. Springer, \bibinfo{pages}{110--126}.
\newblock


\bibitem[Li et~al\mbox{.}(2017)]%
        {li2017universal}
\bibfield{author}{\bibinfo{person}{Yijun Li}, \bibinfo{person}{Chen Fang}, \bibinfo{person}{Jimei Yang}, \bibinfo{person}{Zhaowen Wang}, \bibinfo{person}{Xin Lu}, {and} \bibinfo{person}{Ming-Hsuan Yang}.} \bibinfo{year}{2017}\natexlab{}.
\newblock \showarticletitle{Universal style transfer via feature transforms}.
\newblock \bibinfo{journal}{\emph{Advances in Neural Information Processing Systems}}  \bibinfo{volume}{30} (\bibinfo{year}{2017}).
\newblock


\bibitem[Li et~al\mbox{.}(2022)]%
        {li2022eliminating}
\bibfield{author}{\bibinfo{person}{Zekun Li}, \bibinfo{person}{Zhengyang Geng}, \bibinfo{person}{Zhao Kang}, \bibinfo{person}{Wenyu Chen}, {and} \bibinfo{person}{Yibo Yang}.} \bibinfo{year}{2022}\natexlab{}.
\newblock \showarticletitle{Eliminating gradient conflict in reference-based line-art colorization}. In \bibinfo{booktitle}{\emph{European Conference on Computer Vision}}. Springer, \bibinfo{pages}{579--596}.
\newblock


\bibitem[Liao et~al\mbox{.}(2017)]%
        {liao2017visual}
\bibfield{author}{\bibinfo{person}{Jing Liao}, \bibinfo{person}{Yuan Yao}, \bibinfo{person}{Lu Yuan}, \bibinfo{person}{Gang Hua}, {and} \bibinfo{person}{Sing~Bing Kang}.} \bibinfo{year}{2017}\natexlab{}.
\newblock \showarticletitle{Visual attribute transfer through deep image analogy}.
\newblock \bibinfo{journal}{\emph{arXiv preprint arXiv:1705.01088}} (\bibinfo{year}{2017}).
\newblock


\bibitem[Lin et~al\mbox{.}(2025)]%
        {lin2025sell}
\bibfield{author}{\bibinfo{person}{Jianghao Lin}, \bibinfo{person}{Peng Du}, \bibinfo{person}{Jiaqi Liu}, \bibinfo{person}{Weite Li}, \bibinfo{person}{Yong Yu}, \bibinfo{person}{Weinan Zhang}, {and} \bibinfo{person}{Yang Cao}.} \bibinfo{year}{2025}\natexlab{}.
\newblock \showarticletitle{Sell It Before You Make It: Revolutionizing E-Commerce with Personalized AI-Generated Items}.
\newblock \bibinfo{journal}{\emph{arXiv preprint arXiv:2503.22182}} (\bibinfo{year}{2025}).
\newblock


\bibitem[Liu et~al\mbox{.}(2024)]%
        {liu2024towards}
\bibfield{author}{\bibinfo{person}{Bingyan Liu}, \bibinfo{person}{Chengyu Wang}, \bibinfo{person}{Tingfeng Cao}, \bibinfo{person}{Kui Jia}, {and} \bibinfo{person}{Jun Huang}.} \bibinfo{year}{2024}\natexlab{}.
\newblock \showarticletitle{Towards Understanding Cross and Self-Attention in Stable Diffusion for Text-Guided Image Editing}. In \bibinfo{booktitle}{\emph{Proceedings of the IEEE/CVF Conference on Computer Vision and Pattern Recognition}}. \bibinfo{pages}{7817--7826}.
\newblock


\bibitem[Ma and Sun(2022)]%
        {ma2022multilayered}
\bibfield{author}{\bibinfo{person}{Chen Ma} {and} \bibinfo{person}{Zhengxing Sun}.} \bibinfo{year}{2022}\natexlab{}.
\newblock \showarticletitle{Multilayered stitch generating for random-needle embroidery}.
\newblock \bibinfo{journal}{\emph{The Visual Computer}} \bibinfo{volume}{38}, \bibinfo{number}{11} (\bibinfo{year}{2022}), \bibinfo{pages}{3667--3679}.
\newblock


\bibitem[Mao et~al\mbox{.}(2025)]%
        {mao2025ace++}
\bibfield{author}{\bibinfo{person}{Chaojie Mao}, \bibinfo{person}{Jingfeng Zhang}, \bibinfo{person}{Yulin Pan}, \bibinfo{person}{Zeyinzi Jiang}, \bibinfo{person}{Zhen Han}, \bibinfo{person}{Yu Liu}, {and} \bibinfo{person}{Jingren Zhou}.} \bibinfo{year}{2025}\natexlab{}.
\newblock \showarticletitle{ACE++: Instruction-Based Image Creation and Editing via Context-Aware Content Filling}.
\newblock \bibinfo{journal}{\emph{arXiv preprint arXiv:2501.02487}} (\bibinfo{year}{2025}).
\newblock


\bibitem[Meng et~al\mbox{.}(2021)]%
        {meng2021sdedit}
\bibfield{author}{\bibinfo{person}{Chenlin Meng}, \bibinfo{person}{Yutong He}, \bibinfo{person}{Yang Song}, \bibinfo{person}{Jiaming Song}, \bibinfo{person}{Jiajun Wu}, \bibinfo{person}{Jun-Yan Zhu}, {and} \bibinfo{person}{Stefano Ermon}.} \bibinfo{year}{2021}\natexlab{}.
\newblock \showarticletitle{Sdedit: Guided image synthesis and editing with stochastic differential equations}.
\newblock \bibinfo{journal}{\emph{arXiv preprint arXiv:2108.01073}} (\bibinfo{year}{2021}).
\newblock


\bibitem[Mokady et~al\mbox{.}(2023)]%
        {mokady2023null}
\bibfield{author}{\bibinfo{person}{Ron Mokady}, \bibinfo{person}{Amir Hertz}, \bibinfo{person}{Kfir Aberman}, \bibinfo{person}{Yael Pritch}, {and} \bibinfo{person}{Daniel Cohen-Or}.} \bibinfo{year}{2023}\natexlab{}.
\newblock \showarticletitle{Null-text inversion for editing real images using guided diffusion models}. In \bibinfo{booktitle}{\emph{Proceedings of the IEEE/CVF Conference on Computer Vision and Pattern Recognition}}. \bibinfo{pages}{6038--6047}.
\newblock


\bibitem[Mou et~al\mbox{.}(2024)]%
        {mou2024t2i}
\bibfield{author}{\bibinfo{person}{Chong Mou}, \bibinfo{person}{Xintao Wang}, \bibinfo{person}{Liangbin Xie}, \bibinfo{person}{Yanze Wu}, \bibinfo{person}{Jian Zhang}, \bibinfo{person}{Zhongang Qi}, {and} \bibinfo{person}{Ying Shan}.} \bibinfo{year}{2024}\natexlab{}.
\newblock \showarticletitle{T2i-adapter: Learning adapters to dig out more controllable ability for text-to-image diffusion models}. In \bibinfo{booktitle}{\emph{Proceedings of the AAAI Conference on Artificial Intelligence}}, Vol.~\bibinfo{volume}{38}. \bibinfo{pages}{4296--4304}.
\newblock


\bibitem[Nichol et~al\mbox{.}(2021)]%
        {nichol2021glide}
\bibfield{author}{\bibinfo{person}{Alex Nichol}, \bibinfo{person}{Prafulla Dhariwal}, \bibinfo{person}{Aditya Ramesh}, \bibinfo{person}{Pranav Shyam}, \bibinfo{person}{Pamela Mishkin}, \bibinfo{person}{Bob McGrew}, \bibinfo{person}{Ilya Sutskever}, {and} \bibinfo{person}{Mark Chen}.} \bibinfo{year}{2021}\natexlab{}.
\newblock \showarticletitle{Glide: Towards photorealistic image generation and editing with text-guided diffusion models}.
\newblock \bibinfo{journal}{\emph{arXiv preprint arXiv:2112.10741}} (\bibinfo{year}{2021}).
\newblock


\bibitem[Nichols(2012)]%
        {nichols2012encyclopedia}
\bibfield{author}{\bibinfo{person}{Marion Nichols}.} \bibinfo{year}{2012}\natexlab{}.
\newblock \bibinfo{booktitle}{\emph{Encyclopedia of embroidery stitches, including crewel}}.
\newblock \bibinfo{publisher}{Courier Corporation}.
\newblock


\bibitem[Park and Lee(2019)]%
        {park2019arbitrary}
\bibfield{author}{\bibinfo{person}{Dae~Young Park} {and} \bibinfo{person}{Kwang~Hee Lee}.} \bibinfo{year}{2019}\natexlab{}.
\newblock \showarticletitle{Arbitrary style transfer with style-attentional networks}. In \bibinfo{booktitle}{\emph{proceedings of the IEEE/CVF Conference on Computer Vision and Pattern Recognition}}. \bibinfo{pages}{5880--5888}.
\newblock


\bibitem[Park et~al\mbox{.}(2020)]%
        {park2020contrastive}
\bibfield{author}{\bibinfo{person}{Taesung Park}, \bibinfo{person}{Alexei~A Efros}, \bibinfo{person}{Richard Zhang}, {and} \bibinfo{person}{Jun-Yan Zhu}.} \bibinfo{year}{2020}\natexlab{}.
\newblock \showarticletitle{Contrastive learning for unpaired image-to-image translation}. In \bibinfo{booktitle}{\emph{Computer Vision--ECCV 2020: 16th European Conference, Glasgow, UK, August 23--28, 2020, Proceedings, Part IX 16}}. Springer, \bibinfo{pages}{319--345}.
\newblock


\bibitem[Peebles and Xie(2023)]%
        {peebles2023scalable}
\bibfield{author}{\bibinfo{person}{William Peebles} {and} \bibinfo{person}{Saining Xie}.} \bibinfo{year}{2023}\natexlab{}.
\newblock \showarticletitle{Scalable diffusion models with transformers}. In \bibinfo{booktitle}{\emph{Proceedings of the IEEE/CVF International Conference on Computer Vision}}. \bibinfo{pages}{4195--4205}.
\newblock


\bibitem[Pile(2018)]%
        {pile2018fashion}
\bibfield{author}{\bibinfo{person}{Jessica Pile}.} \bibinfo{year}{2018}\natexlab{}.
\newblock \bibinfo{booktitle}{\emph{Fashion Embroidery: Embroidery Techniques and Inspiration for Haute-Couture Clothing}}.
\newblock \bibinfo{publisher}{Batsford Books}.
\newblock


\bibitem[Podell et~al\mbox{.}(2023)]%
        {podell2023sdxl}
\bibfield{author}{\bibinfo{person}{Dustin Podell}, \bibinfo{person}{Zion English}, \bibinfo{person}{Kyle Lacey}, \bibinfo{person}{Andreas Blattmann}, \bibinfo{person}{Tim Dockhorn}, \bibinfo{person}{Jonas M{\"u}ller}, \bibinfo{person}{Joe Penna}, {and} \bibinfo{person}{Robin Rombach}.} \bibinfo{year}{2023}\natexlab{}.
\newblock \showarticletitle{Sdxl: Improving latent diffusion models for high-resolution image synthesis}.
\newblock \bibinfo{journal}{\emph{arXiv preprint arXiv:2307.01952}} (\bibinfo{year}{2023}).
\newblock


\bibitem[Qi et~al\mbox{.}(2024)]%
        {qi2024deadiff}
\bibfield{author}{\bibinfo{person}{Tianhao Qi}, \bibinfo{person}{Shancheng Fang}, \bibinfo{person}{Yanze Wu}, \bibinfo{person}{Hongtao Xie}, \bibinfo{person}{Jiawei Liu}, \bibinfo{person}{Lang Chen}, \bibinfo{person}{Qian He}, {and} \bibinfo{person}{Yongdong Zhang}.} \bibinfo{year}{2024}\natexlab{}.
\newblock \showarticletitle{DEADiff: An Efficient Stylization Diffusion Model with Disentangled Representations}. In \bibinfo{booktitle}{\emph{Proceedings of the IEEE/CVF Conference on Computer Vision and Pattern Recognition}}. \bibinfo{pages}{8693--8702}.
\newblock


\bibitem[Radford et~al\mbox{.}(2021)]%
        {Radford2021LearningTV}
\bibfield{author}{\bibinfo{person}{Alec Radford}, \bibinfo{person}{Jong~Wook Kim}, \bibinfo{person}{Chris Hallacy}, \bibinfo{person}{A. Ramesh}, \bibinfo{person}{Gabriel Goh}, \bibinfo{person}{Sandhini Agarwal}, \bibinfo{person}{Girish Sastry}, \bibinfo{person}{Amanda Askell}, \bibinfo{person}{Pamela Mishkin}, \bibinfo{person}{Jack Clark}, \bibinfo{person}{Gretchen Krueger}, {and} \bibinfo{person}{Ilya Sutskever}.} \bibinfo{year}{2021}\natexlab{}.
\newblock \showarticletitle{Learning Transferable Visual Models From Natural Language Supervision}. In \bibinfo{booktitle}{\emph{ICML}}. \bibinfo{pages}{8748--8763}.
\newblock


\bibitem[Rombach et~al\mbox{.}(2022)]%
        {rombach2022high}
\bibfield{author}{\bibinfo{person}{Robin Rombach}, \bibinfo{person}{Andreas Blattmann}, \bibinfo{person}{Dominik Lorenz}, \bibinfo{person}{Patrick Esser}, {and} \bibinfo{person}{Bj{\"o}rn Ommer}.} \bibinfo{year}{2022}\natexlab{}.
\newblock \showarticletitle{High-resolution image synthesis with latent diffusion models}. In \bibinfo{booktitle}{\emph{Proceedings of the IEEE/CVF Conference on Computer Vision and Pattern Recognition}}. \bibinfo{pages}{10684--10695}.
\newblock


\bibitem[Rout et~al\mbox{.}(2024)]%
        {rout2024rb}
\bibfield{author}{\bibinfo{person}{Litu Rout}, \bibinfo{person}{Yujia Chen}, \bibinfo{person}{Nataniel Ruiz}, \bibinfo{person}{Abhishek Kumar}, \bibinfo{person}{Constantine Caramanis}, \bibinfo{person}{Sanjay Shakkottai}, {and} \bibinfo{person}{Wen-Sheng Chu}.} \bibinfo{year}{2024}\natexlab{}.
\newblock \showarticletitle{RB-Modulation: Training-Free Personalization of Diffusion Models using Stochastic Optimal Control}.
\newblock \bibinfo{journal}{\emph{arXiv preprint arXiv:2405.17401}} (\bibinfo{year}{2024}).
\newblock


\bibitem[Ruiz et~al\mbox{.}(2023)]%
        {ruiz2023dreambooth}
\bibfield{author}{\bibinfo{person}{Nataniel Ruiz}, \bibinfo{person}{Yuanzhen Li}, \bibinfo{person}{Varun Jampani}, \bibinfo{person}{Yael Pritch}, \bibinfo{person}{Michael Rubinstein}, {and} \bibinfo{person}{Kfir Aberman}.} \bibinfo{year}{2023}\natexlab{}.
\newblock \showarticletitle{Dreambooth: Fine tuning text-to-image diffusion models for subject-driven generation}. In \bibinfo{booktitle}{\emph{Proceedings of the IEEE/CVF Conference on Computer Vision and Pattern Recognition}}. \bibinfo{pages}{22500--22510}.
\newblock


\bibitem[Ryu(2022)]%
        {lorarepo}
\bibfield{author}{\bibinfo{person}{Simo Ryu}.} \bibinfo{year}{2022}\natexlab{}.
\newblock \bibinfo{booktitle}{\emph{Low-rank adaptation for fast text-to-image diffusion fine-tuning}}.
\newblock
\urldef\tempurl%
\url{https://github.com/cloneofsimo/lora}
\showURL{%
\tempurl}


\bibitem[Shah et~al\mbox{.}(2025)]%
        {shah2025ziplora}
\bibfield{author}{\bibinfo{person}{Viraj Shah}, \bibinfo{person}{Nataniel Ruiz}, \bibinfo{person}{Forrester Cole}, \bibinfo{person}{Erika Lu}, \bibinfo{person}{Svetlana Lazebnik}, \bibinfo{person}{Yuanzhen Li}, {and} \bibinfo{person}{Varun Jampani}.} \bibinfo{year}{2025}\natexlab{}.
\newblock \showarticletitle{Ziplora: Any subject in any style by effectively merging loras}. In \bibinfo{booktitle}{\emph{European Conference on Computer Vision}}. Springer, \bibinfo{pages}{422--438}.
\newblock


\bibitem[Shen et~al\mbox{.}(2017)]%
        {shen2017illumination}
\bibfield{author}{\bibinfo{person}{Qiqi Shen}, \bibinfo{person}{Dele Cui}, \bibinfo{person}{Yun Sheng}, {and} \bibinfo{person}{Guixu Zhang}.} \bibinfo{year}{2017}\natexlab{}.
\newblock \showarticletitle{Illumination-preserving embroidery simulation for non-photorealistic rendering}. In \bibinfo{booktitle}{\emph{MultiMedia Modeling: 23rd International Conference, MMM 2017, Reykjavik, Iceland, January 4-6, 2017, Proceedings, Part II 23}}. Springer, \bibinfo{pages}{233--244}.
\newblock


\bibitem[SmilingWolf(2023)]%
        {smilingwolf2023wdconvnexttagger}
\bibfield{author}{\bibinfo{person}{SmilingWolf}.} \bibinfo{year}{2023}\natexlab{}.
\newblock \bibinfo{title}{wd-convnext-tagger-v3}.
\newblock \bibinfo{howpublished}{\url{https://huggingface.co/SmilingWolf/wd-convnext-tagger-v3}}.
\newblock


\bibitem[Sohl-Dickstein et~al\mbox{.}(2015)]%
        {sohl2015deep}
\bibfield{author}{\bibinfo{person}{Jascha Sohl-Dickstein}, \bibinfo{person}{Eric Weiss}, \bibinfo{person}{Niru Maheswaranathan}, {and} \bibinfo{person}{Surya Ganguli}.} \bibinfo{year}{2015}\natexlab{}.
\newblock \showarticletitle{Deep unsupervised learning using nonequilibrium thermodynamics}. In \bibinfo{booktitle}{\emph{International Conference on Machine Learning}}. PMLR, \bibinfo{pages}{2256--2265}.
\newblock


\bibitem[Somepalli et~al\mbox{.}(2024)]%
        {somepalli2024measuring}
\bibfield{author}{\bibinfo{person}{Gowthami Somepalli}, \bibinfo{person}{Anubhav Gupta}, \bibinfo{person}{Kamal Gupta}, \bibinfo{person}{Shramay Palta}, \bibinfo{person}{Micah Goldblum}, \bibinfo{person}{Jonas Geiping}, \bibinfo{person}{Abhinav Shrivastava}, {and} \bibinfo{person}{Tom Goldstein}.} \bibinfo{year}{2024}\natexlab{}.
\newblock \showarticletitle{Measuring Style Similarity in Diffusion Models}.
\newblock \bibinfo{journal}{\emph{arXiv preprint arXiv:2404.01292}} (\bibinfo{year}{2024}).
\newblock


\bibitem[Song et~al\mbox{.}(2020)]%
        {song2020denoising}
\bibfield{author}{\bibinfo{person}{Jiaming Song}, \bibinfo{person}{Chenlin Meng}, {and} \bibinfo{person}{Stefano Ermon}.} \bibinfo{year}{2020}\natexlab{}.
\newblock \showarticletitle{Denoising diffusion implicit models}.
\newblock \bibinfo{journal}{\emph{arXiv preprint arXiv:2010.02502}} (\bibinfo{year}{2020}).
\newblock


\bibitem[{\v{S}}ubrtov{\'a} et~al\mbox{.}(2023)]%
        {vsubrtova2023diffusion}
\bibfield{author}{\bibinfo{person}{Ad{\'e}la {\v{S}}ubrtov{\'a}}, \bibinfo{person}{Michal Luk{\'a}{\v{c}}}, \bibinfo{person}{Jan {\v{C}}ech}, \bibinfo{person}{David Futschik}, \bibinfo{person}{Eli Shechtman}, {and} \bibinfo{person}{Daniel S{\`y}kora}.} \bibinfo{year}{2023}\natexlab{}.
\newblock \showarticletitle{Diffusion image analogies}. In \bibinfo{booktitle}{\emph{ACM SIGGRAPH 2023 Conference Proceedings}}. \bibinfo{pages}{1--10}.
\newblock


\bibitem[Tang et~al\mbox{.}(2024)]%
        {tang2024realfill}
\bibfield{author}{\bibinfo{person}{Luming Tang}, \bibinfo{person}{Nataniel Ruiz}, \bibinfo{person}{Qinghao Chu}, \bibinfo{person}{Yuanzhen Li}, \bibinfo{person}{Aleksander Holynski}, \bibinfo{person}{David~E Jacobs}, \bibinfo{person}{Bharath Hariharan}, \bibinfo{person}{Yael Pritch}, \bibinfo{person}{Neal Wadhwa}, \bibinfo{person}{Kfir Aberman}, {et~al\mbox{.}}} \bibinfo{year}{2024}\natexlab{}.
\newblock \showarticletitle{Realfill: Reference-driven generation for authentic image completion}.
\newblock \bibinfo{journal}{\emph{ACM Transactions on Graphics (TOG)}} \bibinfo{volume}{43}, \bibinfo{number}{4} (\bibinfo{year}{2024}), \bibinfo{pages}{1--12}.
\newblock


\bibitem[Tenenbaum and Freeman(1996)]%
        {tenenbaum1996separating}
\bibfield{author}{\bibinfo{person}{Joshua Tenenbaum} {and} \bibinfo{person}{William Freeman}.} \bibinfo{year}{1996}\natexlab{}.
\newblock \showarticletitle{Separating style and content}.
\newblock \bibinfo{journal}{\emph{Advances in Neural Information Processing Systems}}  \bibinfo{volume}{9} (\bibinfo{year}{1996}).
\newblock


\bibitem[Tewel et~al\mbox{.}(2023)]%
        {tewel2023key}
\bibfield{author}{\bibinfo{person}{Yoad Tewel}, \bibinfo{person}{Rinon Gal}, \bibinfo{person}{Gal Chechik}, {and} \bibinfo{person}{Yuval Atzmon}.} \bibinfo{year}{2023}\natexlab{}.
\newblock \showarticletitle{Key-locked rank one editing for text-to-image personalization}. In \bibinfo{booktitle}{\emph{ACM SIGGRAPH 2023 Conference Proceedings}}. \bibinfo{pages}{1--11}.
\newblock


\bibitem[Tumanyan et~al\mbox{.}(2022)]%
        {tumanyan2022splicing}
\bibfield{author}{\bibinfo{person}{Narek Tumanyan}, \bibinfo{person}{Omer Bar-Tal}, \bibinfo{person}{Shai Bagon}, {and} \bibinfo{person}{Tali Dekel}.} \bibinfo{year}{2022}\natexlab{}.
\newblock \showarticletitle{Splicing vit features for semantic appearance transfer}. In \bibinfo{booktitle}{\emph{Proceedings of the IEEE/CVF Conference on Computer Vision and Pattern Recognition}}. \bibinfo{pages}{10748--10757}.
\newblock


\bibitem[Tumanyan et~al\mbox{.}(2023)]%
        {tumanyan2023plug}
\bibfield{author}{\bibinfo{person}{Narek Tumanyan}, \bibinfo{person}{Michal Geyer}, \bibinfo{person}{Shai Bagon}, {and} \bibinfo{person}{Tali Dekel}.} \bibinfo{year}{2023}\natexlab{}.
\newblock \showarticletitle{Plug-and-play diffusion features for text-driven image-to-image translation}. In \bibinfo{booktitle}{\emph{Proceedings of the IEEE/CVF Conference on Computer Vision and Pattern Recognition}}. \bibinfo{pages}{1921--1930}.
\newblock


\bibitem[Valevski et~al\mbox{.}(2023)]%
        {valevski2023unitune}
\bibfield{author}{\bibinfo{person}{Dani Valevski}, \bibinfo{person}{Matan Kalman}, \bibinfo{person}{Eyal Molad}, \bibinfo{person}{Eyal Segalis}, \bibinfo{person}{Yossi Matias}, {and} \bibinfo{person}{Yaniv Leviathan}.} \bibinfo{year}{2023}\natexlab{}.
\newblock \showarticletitle{Unitune: Text-driven image editing by fine tuning a diffusion model on a single image}.
\newblock \bibinfo{journal}{\emph{ACM Transactions on Graphics (TOG)}} \bibinfo{volume}{42}, \bibinfo{number}{4} (\bibinfo{year}{2023}), \bibinfo{pages}{1--10}.
\newblock


\bibitem[Wang et~al\mbox{.}(2024)]%
        {wang2024instantstyle}
\bibfield{author}{\bibinfo{person}{Haofan Wang}, \bibinfo{person}{Matteo Spinelli}, \bibinfo{person}{Qixun Wang}, \bibinfo{person}{Xu Bai}, \bibinfo{person}{Zekui Qin}, {and} \bibinfo{person}{Anthony Chen}.} \bibinfo{year}{2024}\natexlab{}.
\newblock \showarticletitle{Instantstyle: Free lunch towards style-preserving in text-to-image generation}.
\newblock \bibinfo{journal}{\emph{arXiv preprint arXiv:2404.02733}} (\bibinfo{year}{2024}).
\newblock


\bibitem[Wang et~al\mbox{.}(2023)]%
        {wang2023stylediffusion}
\bibfield{author}{\bibinfo{person}{Zhizhong Wang}, \bibinfo{person}{Lei Zhao}, {and} \bibinfo{person}{Wei Xing}.} \bibinfo{year}{2023}\natexlab{}.
\newblock \showarticletitle{Stylediffusion: Controllable disentangled style transfer via diffusion models}. In \bibinfo{booktitle}{\emph{Proceedings of the IEEE/CVF International Conference on Computer Vision}}. \bibinfo{pages}{7677--7689}.
\newblock


\bibitem[Wu et~al\mbox{.}(2021)]%
        {wu2021styleformer}
\bibfield{author}{\bibinfo{person}{Xiaolei Wu}, \bibinfo{person}{Zhihao Hu}, \bibinfo{person}{Lu Sheng}, {and} \bibinfo{person}{Dong Xu}.} \bibinfo{year}{2021}\natexlab{}.
\newblock \showarticletitle{Styleformer: Real-time arbitrary style transfer via parametric style composition}. In \bibinfo{booktitle}{\emph{Proceedings of the IEEE/CVF International Conference on Computer Vision}}. \bibinfo{pages}{14618--14627}.
\newblock


\bibitem[Xie and Tu(2015)]%
        {xie2015holistically}
\bibfield{author}{\bibinfo{person}{Saining Xie} {and} \bibinfo{person}{Zhuowen Tu}.} \bibinfo{year}{2015}\natexlab{}.
\newblock \showarticletitle{Holistically-nested edge detection}. In \bibinfo{booktitle}{\emph{Proceedings of the IEEE international conference on computer vision}}. \bibinfo{pages}{1395--1403}.
\newblock


\bibitem[Xing et~al\mbox{.}(2024)]%
        {xing2024csgo}
\bibfield{author}{\bibinfo{person}{Peng Xing}, \bibinfo{person}{Haofan Wang}, \bibinfo{person}{Yanpeng Sun}, \bibinfo{person}{Qixun Wang}, \bibinfo{person}{Xu Bai}, \bibinfo{person}{Hao Ai}, \bibinfo{person}{Renyuan Huang}, {and} \bibinfo{person}{Zechao Li}.} \bibinfo{year}{2024}\natexlab{}.
\newblock \showarticletitle{Csgo: Content-style composition in text-to-image generation}.
\newblock \bibinfo{journal}{\emph{arXiv preprint arXiv:2408.16766}} (\bibinfo{year}{2024}).
\newblock


\bibitem[Yan et~al\mbox{.}(2025)]%
        {yan2025image}
\bibfield{author}{\bibinfo{person}{Dingkun Yan}, \bibinfo{person}{Xinrui Wang}, \bibinfo{person}{Zhuoru Li}, \bibinfo{person}{Suguru Saito}, \bibinfo{person}{Yusuke Iwasawa}, \bibinfo{person}{Yutaka Matsuo}, {and} \bibinfo{person}{Jiaxian Guo}.} \bibinfo{year}{2025}\natexlab{}.
\newblock \showarticletitle{Image Referenced Sketch Colorization Based on Animation Creation Workflow}.
\newblock \bibinfo{journal}{\emph{arXiv preprint arXiv:2502.19937}} (\bibinfo{year}{2025}).
\newblock


\bibitem[Yang et~al\mbox{.}(2022)]%
        {yang2022unsupervised}
\bibfield{author}{\bibinfo{person}{Chen Yang}, \bibinfo{person}{Xinrong Hu}, \bibinfo{person}{Yangjun Ou}, \bibinfo{person}{Saishang Zhong}, \bibinfo{person}{Tao Peng}, \bibinfo{person}{Lei Zhu}, \bibinfo{person}{Ping Li}, {and} \bibinfo{person}{Bin Sheng}.} \bibinfo{year}{2022}\natexlab{}.
\newblock \showarticletitle{Unsupervised Embroidery Generation Using Embroidery Channel Attention}. In \bibinfo{booktitle}{\emph{Proceedings of the 18th ACM SIGGRAPH International Conference on Virtual-Reality Continuum and its Applications in Industry}}. \bibinfo{pages}{1--8}.
\newblock


\bibitem[Yang and Sun(2018)]%
        {yang2018paint}
\bibfield{author}{\bibinfo{person}{Kewei Yang} {and} \bibinfo{person}{Zhengxing Sun}.} \bibinfo{year}{2018}\natexlab{}.
\newblock \showarticletitle{Paint with stitches: a style definition and image-based rendering method for random-needle embroidery}.
\newblock \bibinfo{journal}{\emph{Multimedia Tools and Applications}}  \bibinfo{volume}{77} (\bibinfo{year}{2018}), \bibinfo{pages}{12259--12292}.
\newblock


\bibitem[Yang et~al\mbox{.}(2012)]%
        {yang2012image}
\bibfield{author}{\bibinfo{person}{Kewei Yang}, \bibinfo{person}{Jie Zhou}, \bibinfo{person}{Zhengxing Sun}, {and} \bibinfo{person}{Yi Li}.} \bibinfo{year}{2012}\natexlab{}.
\newblock \showarticletitle{Image-based irregular needling embroidery rendering}. In \bibinfo{booktitle}{\emph{proceedings of the 5th International Symposium on Visual Information Communication and Interaction}}. \bibinfo{pages}{87--94}.
\newblock


\bibitem[Yang et~al\mbox{.}(2023)]%
        {yang2023zero}
\bibfield{author}{\bibinfo{person}{Serin Yang}, \bibinfo{person}{Hyunmin Hwang}, {and} \bibinfo{person}{Jong~Chul Ye}.} \bibinfo{year}{2023}\natexlab{}.
\newblock \showarticletitle{Zero-shot contrastive loss for text-guided diffusion image style transfer}. In \bibinfo{booktitle}{\emph{Proceedings of the IEEE/CVF International Conference on Computer Vision}}. \bibinfo{pages}{22873--22882}.
\newblock


\bibitem[Yang et~al\mbox{.}(2025)]%
        {yang2025omnisvg}
\bibfield{author}{\bibinfo{person}{Yiying Yang}, \bibinfo{person}{Wei Cheng}, \bibinfo{person}{Sijin Chen}, \bibinfo{person}{Xianfang Zeng}, \bibinfo{person}{Fukun Yin}, \bibinfo{person}{Jiaxu Zhang}, \bibinfo{person}{Liao Wang}, \bibinfo{person}{Gang Yu}, \bibinfo{person}{Xingjun Ma}, {and} \bibinfo{person}{Yu-Gang Jiang}.} \bibinfo{year}{2025}\natexlab{}.
\newblock \showarticletitle{Omnisvg: A unified scalable vector graphics generation model}.
\newblock \bibinfo{journal}{\emph{arXiv preprint arXiv:2504.06263}} (\bibinfo{year}{2025}).
\newblock


\bibitem[Ye et~al\mbox{.}(2023)]%
        {ye2023ip}
\bibfield{author}{\bibinfo{person}{Hu Ye}, \bibinfo{person}{Jun Zhang}, \bibinfo{person}{Sibo Liu}, \bibinfo{person}{Xiao Han}, {and} \bibinfo{person}{Wei Yang}.} \bibinfo{year}{2023}\natexlab{}.
\newblock \showarticletitle{Ip-adapter: Text compatible image prompt adapter for text-to-image diffusion models}.
\newblock \bibinfo{journal}{\emph{arXiv preprint arXiv:2308.06721}} (\bibinfo{year}{2023}).
\newblock


\bibitem[Ye et~al\mbox{.}(2021)]%
        {ye2021towards}
\bibfield{author}{\bibinfo{person}{Jingwen Ye}, \bibinfo{person}{Yixin Ji}, \bibinfo{person}{Jie Song}, \bibinfo{person}{Zunlei Feng}, {and} \bibinfo{person}{Mingli Song}.} \bibinfo{year}{2021}\natexlab{}.
\newblock \showarticletitle{Towards End-to-End Embroidery Style Generation: A Paired Dataset and Benchmark}. In \bibinfo{booktitle}{\emph{Pattern Recognition and Computer Vision: 4th Chinese Conference, PRCV 2021, Beijing, China, October 29--November 1, 2021, Proceedings, Part IV 4}}. Springer, \bibinfo{pages}{201--213}.
\newblock


\bibitem[Zhang et~al\mbox{.}(2023c)]%
        {zhang2023adding}
\bibfield{author}{\bibinfo{person}{Lvmin Zhang}, \bibinfo{person}{Anyi Rao}, {and} \bibinfo{person}{Maneesh Agrawala}.} \bibinfo{year}{2023}\natexlab{c}.
\newblock \showarticletitle{Adding conditional control to text-to-image diffusion models}. In \bibinfo{booktitle}{\emph{Proceedings of the IEEE/CVF International Conference on Computer Vision}}. \bibinfo{pages}{3836--3847}.
\newblock


\bibitem[Zhang et~al\mbox{.}(2021)]%
        {zhang2021line}
\bibfield{author}{\bibinfo{person}{Qian Zhang}, \bibinfo{person}{Bo Wang}, \bibinfo{person}{Wei Wen}, \bibinfo{person}{Hai Li}, {and} \bibinfo{person}{Junhui Liu}.} \bibinfo{year}{2021}\natexlab{}.
\newblock \showarticletitle{Line art correlation matching feature transfer network for automatic animation colorization}. In \bibinfo{booktitle}{\emph{Proceedings of the IEEE/CVF Winter Conference on Applications of Computer Vision}}. \bibinfo{pages}{3872--3881}.
\newblock


\bibitem[Zhang et~al\mbox{.}(2018)]%
        {zhang2018perceptual}
\bibfield{author}{\bibinfo{person}{Richard Zhang}, \bibinfo{person}{Phillip Isola}, \bibinfo{person}{Alexei~A Efros}, \bibinfo{person}{Eli Shechtman}, {and} \bibinfo{person}{Oliver Wang}.} \bibinfo{year}{2018}\natexlab{}.
\newblock \showarticletitle{The Unreasonable Effectiveness of Deep Features as a Perceptual Metric}. In \bibinfo{booktitle}{\emph{CVPR}}. \bibinfo{pages}{586--595}.
\newblock


\bibitem[Zhang et~al\mbox{.}(2023a)]%
        {zhang2023prospect}
\bibfield{author}{\bibinfo{person}{Yuxin Zhang}, \bibinfo{person}{Weiming Dong}, \bibinfo{person}{Fan Tang}, \bibinfo{person}{Nisha Huang}, \bibinfo{person}{Haibin Huang}, \bibinfo{person}{Chongyang Ma}, \bibinfo{person}{Tong-Yee Lee}, \bibinfo{person}{Oliver Deussen}, {and} \bibinfo{person}{Changsheng Xu}.} \bibinfo{year}{2023}\natexlab{a}.
\newblock \showarticletitle{Prospect: Prompt spectrum for attribute-aware personalization of diffusion models}.
\newblock \bibinfo{journal}{\emph{ACM Transactions on Graphics (TOG)}} \bibinfo{volume}{42}, \bibinfo{number}{6} (\bibinfo{year}{2023}), \bibinfo{pages}{1--14}.
\newblock


\bibitem[Zhang et~al\mbox{.}(2023b)]%
        {zhang2023inversion}
\bibfield{author}{\bibinfo{person}{Yuxin Zhang}, \bibinfo{person}{Nisha Huang}, \bibinfo{person}{Fan Tang}, \bibinfo{person}{Haibin Huang}, \bibinfo{person}{Chongyang Ma}, \bibinfo{person}{Weiming Dong}, {and} \bibinfo{person}{Changsheng Xu}.} \bibinfo{year}{2023}\natexlab{b}.
\newblock \showarticletitle{Inversion-based style transfer with diffusion models}. In \bibinfo{booktitle}{\emph{Proceedings of the IEEE/CVF Conference on Computer Vision and Pattern Recognition}}. \bibinfo{pages}{10146--10156}.
\newblock


\bibitem[Zhang et~al\mbox{.}(2022)]%
        {zhang2022domain}
\bibfield{author}{\bibinfo{person}{Yuxin Zhang}, \bibinfo{person}{Fan Tang}, \bibinfo{person}{Weiming Dong}, \bibinfo{person}{Haibin Huang}, \bibinfo{person}{Chongyang Ma}, \bibinfo{person}{Tong-Yee Lee}, {and} \bibinfo{person}{Changsheng Xu}.} \bibinfo{year}{2022}\natexlab{}.
\newblock \showarticletitle{Domain enhanced arbitrary image style transfer via contrastive learning}. In \bibinfo{booktitle}{\emph{ACM SIGGRAPH 2022 Conference Proceedings}}. \bibinfo{pages}{1--8}.
\newblock


\bibitem[Zhenyuan et~al\mbox{.}(2023)]%
        {zhenyuan2023embroidery}
\bibfield{author}{\bibinfo{person}{Liu Zhenyuan}, \bibinfo{person}{Michal Piovar\u{c}i}, \bibinfo{person}{Christian Hafner}, \bibinfo{person}{Rapha\"{e}l Charrondi\`{e}re}, {and} \bibinfo{person}{Bernd Bickel}.} \bibinfo{year}{2023}\natexlab{}.
\newblock \showarticletitle{Directionality-Aware Design of Embroidery Patterns}.
\newblock \bibinfo{journal}{\emph{Computer Graphics Forum}} \bibinfo{volume}{42}, \bibinfo{number}{2} (\bibinfo{year}{2023}).
\newblock
\urldef\tempurl%
\url{https://doi.org/10.1111/cgf.14770}
\showDOI{\tempurl}


\bibitem[Zhou et~al\mbox{.}(2025)]%
        {zhou2025attention}
\bibfield{author}{\bibinfo{person}{Yang Zhou}, \bibinfo{person}{Xu Gao}, \bibinfo{person}{Zichong Chen}, {and} \bibinfo{person}{Hui Huang}.} \bibinfo{year}{2025}\natexlab{}.
\newblock \showarticletitle{Attention distillation: A unified approach to visual characteristics transfer}.
\newblock \bibinfo{journal}{\emph{arXiv preprint arXiv:2502.20235}} (\bibinfo{year}{2025}).
\newblock


\bibitem[Zhu et~al\mbox{.}(2017)]%
        {zhu2017unpaired}
\bibfield{author}{\bibinfo{person}{Jun-Yan Zhu}, \bibinfo{person}{Taesung Park}, \bibinfo{person}{Phillip Isola}, {and} \bibinfo{person}{Alexei~A Efros}.} \bibinfo{year}{2017}\natexlab{}.
\newblock \showarticletitle{Unpaired image-to-image translation using cycle-consistent adversarial networks}. In \bibinfo{booktitle}{\emph{Proceedings of the IEEE International Conference on Computer Vision}}. \bibinfo{pages}{2223--2232}.
\newblock


\bibitem[Zou et~al\mbox{.}(2021)]%
        {zou2021stylized}
\bibfield{author}{\bibinfo{person}{Zhengxia Zou}, \bibinfo{person}{Tianyang Shi}, \bibinfo{person}{Shuang Qiu}, \bibinfo{person}{Yi Yuan}, {and} \bibinfo{person}{Zhenwei Shi}.} \bibinfo{year}{2021}\natexlab{}.
\newblock \showarticletitle{Stylized neural painting}. In \bibinfo{booktitle}{\emph{Proceedings of the IEEE/CVF Conference on Computer Vision and Pattern Recognition}}. \bibinfo{pages}{15689--15698}.
\newblock


\end{thebibliography}

\appendix

\end{document}